\documentclass[a4paper,fleqn]{cas-sc}

\usepackage{comment}
\usepackage{graphicx}
\usepackage{bm}

\usepackage[numbers]{natbib}

\usepackage{float}
\makeatletter
\AtBeginDocument{%
  \let\figure\@undefined
  \let\endfigure\@undefined
  \newenvironment{figure}[1][tbp]%
    {\@float{figure}[#1]\sffamily\small\centering}%
    {\end@float}%
}
\makeatother

\def\tsc#1{\csdef{#1}{\textsc{\lowercase{#1}}\xspace}}
\tsc{WGM}
\tsc{QE}
\tsc{EP}
\tsc{PMS}
\tsc{BEC}
\tsc{DE}

\begin{document}
\let\WriteBookmarks\relax
\renewcommand{\floatpagefraction}{0.99}
\renewcommand{\textfraction}{0.01}
\renewcommand{\topfraction}{0.99}
\renewcommand{\bottomfraction}{0.99}

\shorttitle{MSSP}
\title [mode = title]{Simulating Torsional Vibrations of Faulty Bevel Gears Using the Polygonal Contact Method} 






\author[1]{Milla Vehviläinen}[type=editor, auid=000,bioid=1, orcid=0000-0001-9898-9854]
\cormark[1] 
\ead{milla.vehvilainen@vtt.fi} 

\author[2]{Aleksanteri Hämäläinen}[orcid=0009-0008-6470-3617]
\author[1]{Pekka Rahkola}[orcid=0000-0003-3932-9957]
\author[1]{Mikko Savolainen}[orcid=0009-0001-8066-5410]
\author[1]{Janne Keränen}[orcid=0000-0003-1195-0856]
\author[1]{Jari Halme}[orcid=0000-0002-8593-2477]
\author[1]{Jenni Pippuri-Mäkeläinen}[orcid=0000-0002-9698-1549]
\author[2]{Kari Tammi}[orcid=0000-0001-9376-2386]
\author[3]{Anouar Belahcen}[orcid=0000-0003-2154-8692]

\affiliation[1]{
    organization={VTT Technical Research Centre of Finland Ltd.},
    addressline={P.O.\ Box 1000, FI-02044 VTT}, 
    country={Finland}}
\affiliation[2]{
    organization={Aalto University, Department of Mechanical Engineering},
    addressline={FI-02150}, 
    city={Espoo},
    country={Finland}}
\affiliation[3]{
    organization={Aalto University, Department of Electrical Engineering and Automation},
    addressline={FI-00076}, 
    city={Espoo},
    country={Finland}}
    

\cortext[cor1]{Corresponding author}

\nonumnote{This work was funded by the Research Council of Finland through 'The Centre of Excellence in High-Speed Electromechanical Energy Conversion Systems' (HiECSs), decision no. 346441, and by the European Union's NextGenerationEU under the RePowerEU program for sustainable growth in Finland.}

\shortauthors{Vehviläinen et~al.}



\begin{abstract}
    Gears are an integral component of electromechanical applications, but accurate condition monitoring methods, including data-driven predictive maintenance, are strongly dependent on high-quality data, especially from faulty components. To address the scarcity of data, we proposed a multibody simulation using an advanced polygonal contact method to replicate torsional vibrations from an experimental azimuth thruster test rig. The key novelty is the ability to simulate both healthy and faulty gears with arbitrary fault geometries. The simulated signals closely matched the measurements in both time and frequency domains. In the time domain, average torque levels and periodic fluctuations aligned well, although measured signals exhibited higher peak-to-peak amplitudes and greater noise, particularly in healthy conditions at lower rotational speeds. In the frequency domain, the simulations accurately reproduced expected fault frequencies and corresponding sidebands, with larger faults producing higher amplitudes. While the simulations tended to overestimate peak amplitudes and underestimate external noise, the results were highly comparable to measurements and consistent with the physical expectations. These findings provide a robust foundation for enhancing data-driven condition monitoring methods, particularly those employing machine learning or deep learning.
\end{abstract}


\maketitle

\section{Introduction} \label{sec:sect_intro}

Gears are critical components in power transmission systems, enabling torque and speed conversion in various applications, including automotive, maritime, and industrial sectors. Their failure can lead to costly downtimes, safety hazards, and diminished system performances, highlighting the need for their condition monitoring (CM) \cite{talluri2023}. Consequently, CM is vital to ensure reliable and efficient operation, minimise unplanned outages, and optimise maintenance planning \cite{aherwar2012}.

Recently, data-driven predictive maintenance (PdM) has emerged as a strategy for maintaining gear systems using CM data, particularly vibration signals \cite{surucu2023,matania2024,soomro2024}. In this approach, machine learning (ML) and deep learning (DL) algorithms are applied to automatically detect early signs of faults, identify fault types, or estimate the remaining useful life of components. These methods aim to reduce unplanned downtime by enabling proactive maintenance actions. Unlike traditional knowledge-driven methods, which rely on expert-defined rules and thresholds, data-driven techniques learn patterns directly from historical data. For example, ML models can be trained to recognise fault signatures by extracting and comparing features from past cases. DL methods go a step further by learning relevant features directly from raw vibration data, reducing the need for manual preprocessing. However, these algorithms, no matter how advanced, can only deliver reliable results when trained with sufficient and representative data. In general, more high-quality data leads to more accurate predictions.

Gear fault detection has traditionally relied on lateral vibration measurements from accelerometers \cite{surucu2023,matania2024,soomro2024}. While effective, lateral vibrations are often masked by structural resonances, damping, and external noise, making it difficult to isolate fault-specific features. As an alternative, torsional vibrations, observable through variations in rotational speed and torque, can provide a more direct information, particularly regarding disturbances in load transmission and gear meshing due to defects \cite{liu2018,miettinen2023}. Torsional signals are generally less sensitive to non-shaft-related structural resonances and may offer clearer insights at lower frequencies, which aligns with their higher sensitivity to slow dynamic changes \cite{xue2018}. As a result, torsional vibration analysis is emerging as a promising strategy for gear fault detection and PdM, especially in setups where conventional lateral vibration analysis faces limitations.

However, a well-recognised bottleneck in data-driven CM (condition monitoring) is the limited availability of experimental measurements, especially for faulty scenarios. Acquiring such datasets remains particularly challenging due to the rarity of real-world failure cases, the high capital cost and instrumentation complexity of test rigs, and the difficulty in replicating realistic operating conditions and duty cycles in a controlled environment \cite{soto2020}. These constraints create a compelling motivation to use simulation-based approaches to augment and complement experimental datasets, allowing a broader and more systematic fault analysis, and reducing the dependence on costly physical tests \cite{aherwar2012}. Provided that the simulation models are sufficiently accurate and physically representative, simulated data can partially substitute for experimental measurements, especially at operating points that would otherwise be unrealisable.

Gear faults are mechanical phenomena that influence system dynamics through localised changes in the gear tooth contact geometry. Numerical methods, such as multibody simulation (MBS), are well suited for capturing such dynamic behaviour in mechanical systems, including possible defect effects. Conventional dynamic models often assume idealised component shapes, which are appropriate for representing healthy components but insufficient for simulating realistic fault conditions. Instead of directly modelling fault geometries, common strategies include altering secondary parameters to resemble faulty behaviour. For example, fault effects can be approximated by modifying torsional compliance, backlash, or dynamic shifting, particularly under harsh conditions \cite{prokop2018}. Eccentricity can be modelled by displacing the centreline of a gearwheel or exaggerating the phase angle \cite{lafi2019}, while wear and cracks are typically represented via local reductions in stiffness or friction properties \cite{chen2022, koutsoupakis2023, han2023}. Only a few studies have incorporated actual faults into their gear simulations represented by notch geometries or missing teeth \cite{koutsoupakis2023, furch2017, kumar2023, kumar2025}. Even in these cases, the underlying contact algorithms remain anchored in classical Hertzian contact theory, which is mostly applicable to contacts between simple shapes, such as cylinders and spheres, with low impact dynamics \cite{flores2022}. Real fault geometries, however, are often rather arbitrary and non-symmetric.

Polygonal contact method (PCM), originally introduced in \cite{hippmann2004,hippmann2024}, is a geometry-based multipoint contact based on the actual surface geometries of the contacting bodies. The body surfaces are described by polygonal meshes composed of triangular discretisation. PCM accounts for the local material properties and surface topography to represent the contact dynamics with greater fidelity. Unlike the majority of the traditional contact models, PCM does not assume continuous contact, which facilitates the simulation of realistic contact dynamics, such as clearance and slipping.
The practical advantage is the possibility to modify the model geometry, including faults, directly through CAD design. 

PCM has been applied in many contact-critical MBS applications, including the biomedical field \cite{synek2012}, aerospace engineering \cite{schafer2010,zhang2021,lojewski2020}, and transportation technology \cite{hajzman2012,weicheng2025}, among others. More recently, PCM has been adapted to simulate faulty ball bearings operating under relatively high surface velocities, enabling the generation of dynamic response data for training an ML-based fault classifier \cite{vehvilainen2023,vehvilainen2024}. These results demonstrate the capability of PCM to support data availability by facilitating realistic physics-based simulation data for PdM.

In this study, we develop an MBS system-model of a thruster driveline, comprising three shafts (drive, middle, and propeller) with flexible elements, interconnected by bevel gears and driven by electric motors under varying loads and speeds. One of the bevel gears is modelled by using the PCM to establish a more realistic gear meshing, specifically under faulty conditions. The introduced gear fault type is a tooth flank fracture (TFF) at two different levels of severity. We aim to assess the feasibility and accuracy of the PCM for simulating gear contact dynamics for PdM applications. In addition to the component-level gear meshing, our simulation model considers the complete drivetrain, allowing for a system-level analysis that accounts for interactions between multiple components and disturbance sources. Torsional vibrations are simulated across the driveline under both healthy and faulty scenarios.

The validity of the proposed simulation model is assessed by comparing the results against measurements from an existing thruster test rig \cite{haikonen2022}, from which a comprehensive dataset has been acquired \cite{dahl2024}. The dataset includes torsional vibration signals measured by using dedicated sensors in healthy and faulty conditions. The primary contributions of this research are to:

\begin{itemize}
    \item Develop an MBS model with an advanced geometry-based PCM that accurately replicates the experimental test setup.
    \item Validate the simulation model through comparisons with experimental measurements of torsional vibrations and calibration of PCM properties for gear meshing, referencing an established Hertzian-based contact module.
    \item Establish a simulation-based framework for generating high-quality synthetic datasets to support data-driven PdM and fault detection.
\end{itemize} 

The remainder of the article is structured as follows. Section \ref{sec:sect_gears} reviews the gear dynamics and fault detection mechanisms, while Section \ref{sec:sect_pcm} introduces the PCM theory. The validation case is presented in Section \ref{sec:sect_exp}. Sections \ref{sec:sect_sim} and \ref{sec:sect_results} describe the simulation model and discuss the results, respectively. Lastly, Section \ref{sec:sect_futurework} outlines the future framework, and Section \ref{sec:sect_conclusion} concludes.

\section{Gearwheel dynamics} \label{sec:sect_gears}
Gears are primarily designed to modify rotational speed, torque, or the direction of rotation in mechanical systems. Gears transmit power through the meshing of the gear teeth, involving elastic deformation, friction, and combined rolling-sliding interactions. Force distribution during gear meshing depends primarily on gear geometry, material properties, and applied load conditions \cite{guiggiani2008}. Under ideal conditions, gear mesh occurs smoothly across contacting teeth, transmitting torque with minimal fluctuation. However, real-world factors, such as pitting, wear, and cracks, disrupt this process by altering the contact stiffness and increasing vibration levels \cite{shweiki2017}, especially when such faults exceed the compensation provided by lubrication and structural damping. These disturbances manifest as characteristic changes in vibration signals, which can be analysed for CM and fault diagnosis. Practical operating conditions also introduce dynamic phenomena, including torsional oscillations and periodic variations in mesh stiffness, significantly affecting the overall dynamic response of the gear system.

    \subsection{Torsional vibrations}
    Torsional vibrations are angular oscillations around the mean rotational speed in rotating shafts due to periodic variations in torque. These oscillations in gear systems arise from meshing stiffness fluctuations, backlash, and inertial effects \cite{kahraman1999}, as well as from changes in the load torque and speed.

    Analysing torsional vibrations is fundamental when designing large rotating machinery, as they directly affect gear performance and longevity. Compared to lateral signals, torsional vibrations contain less external noise, are more sensitive to variations in meshing stiffness caused by defects, and provide a direct measurement of torque transmission efficiency \cite{xue2018, mahapatra2023}. More broadly, defects in gear meshing stiffness manifest clearly in torsional dynamics, making torsional measurements a valuable diagnostic means \cite{miettinen2023,ding2025}. Recently, an attention has grown toward using torsional signals as a diagnostic tool for CM, enabling early fault detection such as tooth breakage and misalignment \cite{xue2018, moghadam2021}.

    Additionally, torsional vibrations are generally simpler to simulate than lateral vibrations. Torsional models are primarily based on rigid-body dynamics, and while flexibility is sometimes included, it usually applies only to rotating components such as shafts or couplings. In contrast, lateral vibrations are influenced by shaft bending and support compliance, often requiring the modelling of flexible structures like bearings or housings. As torsional shaft dynamics are mainly governed by gear meshing and rotational inertia, the simulations are more straightforward and computationally efficient.

    Torsional vibrations are typically measured by using rotational sensors such as optical encoders, laser Doppler vibrometers, or torque transducers mounted on the shaft. Torque transducers provide the direct measurement of transmitted torque variations and are commonly used in test bench setups. 

    In drivetrain mechanics, torque and rotational speed are subjected to real-world effects such as friction, structural damping, and component compliance, all of which contribute to power losses. These lead to observable deviations in torque transmission and rotational speed along the drivetrain. To quantify these losses, an instantaneous shaft power $P$ can be calculated at different points in the system, and the power loss $P_{loss}$ between two points can then be defined as the difference, as follows:

    \vspace{-\baselineskip}
    \begin{equation} \label{eq:power}
        P = T \cdot \omega,
    \end{equation}    
    \begin{align} \label{eq:power_loss}
        P_{loss} = P_{in} - P_{out},
    \end{align}
    
    \noindent where $T$ and $\omega$ are the torque and angular velocity at a given point in the drivetrain, used to calculate the shaft power either at the input ($P_{in}$) or the output ($P_{out}$) shaft.
    
    \subsection{Characteristic fault frequencies}
    In an ideal scenario with perfectly healthy and correctly installed gears, vibrations primarily occur at the gear mesh frequency and its harmonics. The mesh frequency represents the rate at which gear teeth engage and is determined by multiplying the rotational frequency of the shaft by the number of teeth on the gear. In the presence of faults, such as a broken pinion tooth, or manufacturing or assembly errors, additional vibrations can arise due to periodic changes in the meshing stiffness of the gear pair. These changes repeat at a frequency relative to the rotational frequency of the faulty gear. This results in sidebands around the gear mesh frequency due to amplitude modulation effects, such as load variations and resonance interactions. The periodic impact forces generated by faults introduce frequency components that provide diagnostic information regarding the health of the system. The fundamental gear mesh frequency $f_{GMF}$, its multiples, and the associated sidebands $f_{sb}$ are calculated as in Equations \eqref{eq:f_{GMF}} and \eqref{eq:f_{sb}}, respectively:

    \vspace{-\baselineskip}
    \begin{equation} \label{eq:f_{GMF}}
        f_{GMF} = z_n \cdot f_r,
    \end{equation}
    \begin{align} \label{eq:f_{sb}}
        f_{sb} =  f_{GMF} \pm k \cdot f_{r},
    \end{align}

    \noindent where $z_n$ is the number of teeth on the driving gearwheel that is mounted on the modulating shaft rotating at frequency $f_r$. The harmonic number of the modulation frequency, denoted by $k$, is introduced by faults or irregularities, such as gear tooth damage, shaft misalignment, or bearing defects. The specific fault type influences the frequency content of the vibration signal, enabling fault classification through spectral analysis \cite{zeng2025}.

    \subsection{Data processing}
    Measured vibration signals from real-life applications inevitably contain noise and various disturbances, even under healthy operating conditions. Therefore, effective signal processing methods are crucial for extracting relevant fault-related information. Typically, signals and their features are analysed in either the time or frequency domain.
    
    In the time domain, windowing is employed to minimise data leakage, particularly when preparing signals for frequency-domain analysis. After windowing, the Fast Fourier Transform (FFT) converts the time-domain data into the frequency domain. The FFT is widely used in applications involving constant speeds because it effectively decomposes complex vibration signals into individual frequency components, thus highlighting fault-induced spectral features.
    
    Torsional vibration signals typically contain relevant gear dynamic information, such as shaft rotation, gear meshing, and fault-induced modulations, at relatively low frequencies. Applying a low-pass filter can help eliminate irrelevant high-frequency components, originating from electrical noise, sensor vibrations, or structural resonances which can mask or distort fault-induced patterns, especially in time-domain data.
    
    When comparing measured and simulated signals, it is natural to observe some differences, particularly in the time domain. This arises because simulation models inherently simplify the real system and often exclude certain physical phenomena. Therefore, frequency-domain analysis is beneficial, as it emphasises significant features and peak amplitudes while minimising irrelevant information. Such patterns are typically difficult to identify clearly within raw time-domain signals.
    
    In practice, if a defect-related dynamic feature is clearly observable by human inspection in either measured or simulated data, ML and DL algorithms should also reliably detect it. However, human observers might overlook subtle yet important features that are detectable by these algorithms before faults escalate. This capability underscores the importance of PdM methods, which identify and respond to subtle indicators of incipient faults.

\section{Polygonal contacts} \label{sec:sect_pcm}
Accurate modelling of dynamic systems such as gears relies heavily on precise contact descriptions. Contact between colliding bodies is a nonlinear phenomenon involving normal and tangential forces. The contact description directly affects the accuracy of the simulated output data. Contact forces are typically defined as continuous functions of relative penetration between contacting bodies, governed by spring-damper elements \cite{corral2021review}. The spring component determines the elasticity at the contact point, while the damper component accounts for energy dissipation.  

The Hertzian contact model is the most widely accepted analytical reference for contact analysis, commonly used to simulate smooth, undamaged contacts with simple, continuous geometries, such as cylinders and spheres, with low-impact dynamics \cite{flores2022, bejar2024}. Through analytical equations, it approximates contact interactions as ellipsoidal pressure distributions, with contact area increasing linearly with penetration. Consequently, Hertzian theory is not applicable for complex or asymmetric contact geometries often found in real-world applications, especially in faulty cases. Nevertheless, Hertzian theory provides a well-established foundation and reference point for developing more sophisticated contact-modelling techniques. Thus, in this study, the Hertzian model primarily serves as a validation tool, providing a benchmark for calibrating PCM parameters and verifying the accuracy of simulations in healthy contact scenarios. 

PCM is a geometry-based multipoint contact approach considering the actual surface geometries of the contacting bodies described by polygonal meshes. This high-accuracy contact method can be applied in any open-source platform designed for MBS. Furthermore, Simpack software provides a separate polygonal contact force element called FE199 \cite{simpackmanual2022_fe199}, particularly based on PCM. 

In existing contact methods, the intersection detection phase at a particular timestep and position is of iterative nature and the most intensive operation computationally \cite{flores2022}. The contact detection in PCM, in turn, is based on a bounding volume (BV) hierarchy to reduce computational effort \cite{hippmann2004}. This also enables the contact calculation to be divided into multiple threads that can be executed in parallel, further reducing computation demands. PCM determines contact force in the normal direction by the elastic layer stiffness, which is calculated for each polygon within the contact patch. The stiffness model is a rigid body model which is assumed to be covered by a thin layer of elastic springs -- one for each polygon. Equations (\ref{eq:pcm_elastic_modulus}--\ref{eq:pcm_ck}) introduce the contact description according to \cite{hippmann2004}, and Equation \eqref{eq:pcm_ctot} shows the integration in Simpack \cite{simpackmanual2022_fe199}. With homogeneous compressible materials, the elastic modulus $K$ can be derived as follows:

\vspace{-\baselineskip}
\begin{equation} \label{eq:pcm_elastic_modulus}
    K = \frac{1-\nu}{(1+\nu)(1-2\nu)} \cdot E,
\end{equation}       

\noindent where material properties Young’s modulus $E$ and Poisson’s ratio $\nu$ define linear-elastic characteristics for the layers, constant for each polygon. PCM neglects tangential shear stress in a layer of thickness, which results in a direct relation of the normal penetration and pressure:

\vspace{-\baselineskip}
\begin{equation} \label{eq:pcm_normal_pressure}
    p = \frac{K}{b} \cdot u,
\end{equation}

\noindent where $u$ and $p$ are the normal penetration and the surface pressure, both assumed to be constant for each contact element. The elastic layer thickness is denoted by $b$. The normal force $F_{k}$ in an individual contact element $k$ then results in:

\vspace{-\baselineskip}
\begin{equation} \label{eq:pcm_normal_force}
    F_{k} = c_{k} \cdot A_{k} \cdot u_{k},
\end{equation}
        
\noindent where $A_{k}$ is the area of the contacting polygon and $c_{k}$ is the corresponding contact stiffness. Furthermore, contact stiffness $c_{k}$ for a single polygon $k$ depends on the polygon size and the elastic layer: 

\vspace{-\baselineskip}
\begin{equation} \label{eq:pcm_ck}
    c_{k} = \frac{K}{b} \cdot A_{k}.
\end{equation}

Lastly, the total stiffness $c_{tot}$ of a certain contact patch is determined as a combination of primary and secondary surface stiffness \cite{simpackmanual2022_fe199}:

\vspace{-\baselineskip}
\begin{equation} \label{eq:pcm_ctot}
    c_{tot}(u) = \frac{c_{1} \cdot c_{2}}{c_{1} + c_{2}} \cdot {f(u)},
\end{equation}

\noindent where $c_{1}$ and $c_{2}$ are the normal contact stiffnesses on the primary and secondary polygon, respectively. The total normal stiffness $c_{tot}$ can be further scaled by using an input function $f(u)$ that depends on the normal penetration. The polygon size inherently has a significant impact on contact properties, especially with small penetrations, which is when $f(u)$ can be used to adapt contact response for certain applications. In most practical cases, gears included, a sufficient discretisation makes such tuning unnecessary, so the scale function is typically assumed to equal one.

In addition to normal contact forces, the complete contact model needs to consider tangential friction forces. For this, Coulomb's law is commonly used in gears \cite{furch2017,kumar2025}, often in a regularised form like Barnard's friction, which eliminates discontinuities near zero to velocities. A regularised single friction coefficient value remains constant, irrespective of the shaft rotation speed, thereby eliminating stiction-friction. Tangential friction forces are defined as:

\vspace{-\baselineskip}
\begin{equation} \label{eq:bernard_frict}
    F_t = \begin{cases} \mu \cdot F_n \cdot min(k_{\varepsilon} \cdot v_t, 1) & \mbox{if } v_t \geq 0, \\ 
    \mu \cdot F_n \cdot max(k_{\varepsilon} \cdot v_t, -1) & \mbox{if } v_t < 0, \end{cases}
\end{equation}

\noindent where $\mu$ is the friction coefficient and $F_{n}$ is the normal contact force. The relative tangential velocity of the contacting surfaces is represented as $v_t$, and $k_{\varepsilon}$ represents the slope of the linear transition at low relative tangential velocities. Overall, Coulomb's dry friction model provides a simple yet accurate representation of the frictional behaviour of the contacts, and it is numerically efficient and straightforward to apply in dynamic simulation systems \cite{pennestri2016}.

\section{Experimental test rig for validation} \label{sec:sect_exp}
\begin{figure}[H]
        \centering
        \includegraphics[width=0.7\linewidth, trim={0.1cm 0.1cm 0.1cm 0cm},clip]{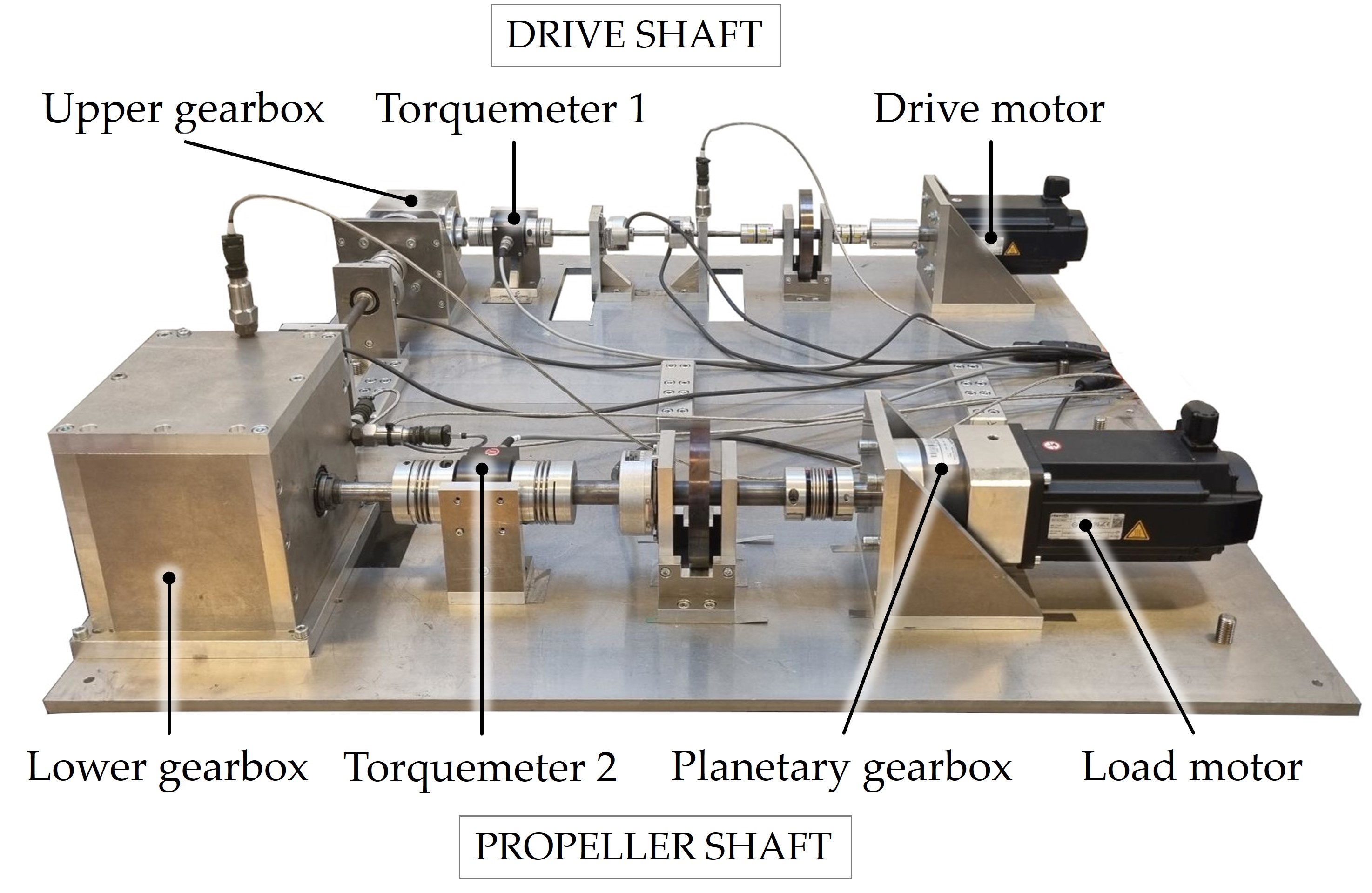}
        \vspace{-4mm}
        \caption{The downscaled test rig of the azimuth thruster with the Z-drive and bevel gears was used for obtaining the experimental data.}
        \label{fig:aalto_gear}
    \end{figure}

Experimental data in this study were obtained from a downscaled bevel gear test bench representing a real azimuth thruster. The test bench was geometrically scaled at 9:1 and fixed in a horizontal position \cite{haikonen2022}. This setup was originally built to generate realistic data under operating conditions comparable to a full-size application. Particular care was taken in appropriately scaling the components to replicate the dynamic characteristics of a full-scale thruster \cite{laine2025a}. Figure \ref{fig:aalto_gear} presents the downscaled thruster test rig. 

    \subsection*{Equipment}
    The test rig features a typical Z-drive layout, comprising three shafts: the drive shaft, middle shaft, and propeller shaft, interconnected with two angular reduction gearboxes with gear ratios of 3:1 and 4:1, respectively. 
    The upper gearbox transmits power from the drive shaft to the middle shaft, while the lower gearbox connects the middle shaft to the propeller shaft. Two identical synchronous servomotors power the system. The drive motor provides propulsion through the drive shaft, and the load motor acts as a propeller, generating torque on the propeller shaft. To achieve increased torque, the load motor is coupled to a planetary gearbox with a gear ratio of 1:8.

    The test rig uses two flywheels to match the lowest natural frequencies of the real counterpart. The first is on the drive shaft between the motor and the upper gearbox. It adds inertia and lowers torsional frequencies, with flexible elastomer couplings to reduce the drive shaft stiffness. The second is on the propeller shaft between the load motor's planetary gearbox and the lower gearbox, simulating the propeller mass.

\begin{figure}[H]
            \centering
            \includegraphics[width=0.52\linewidth, trim={0.4cm 0.4cm 0.4cm 0.4cm},clip]{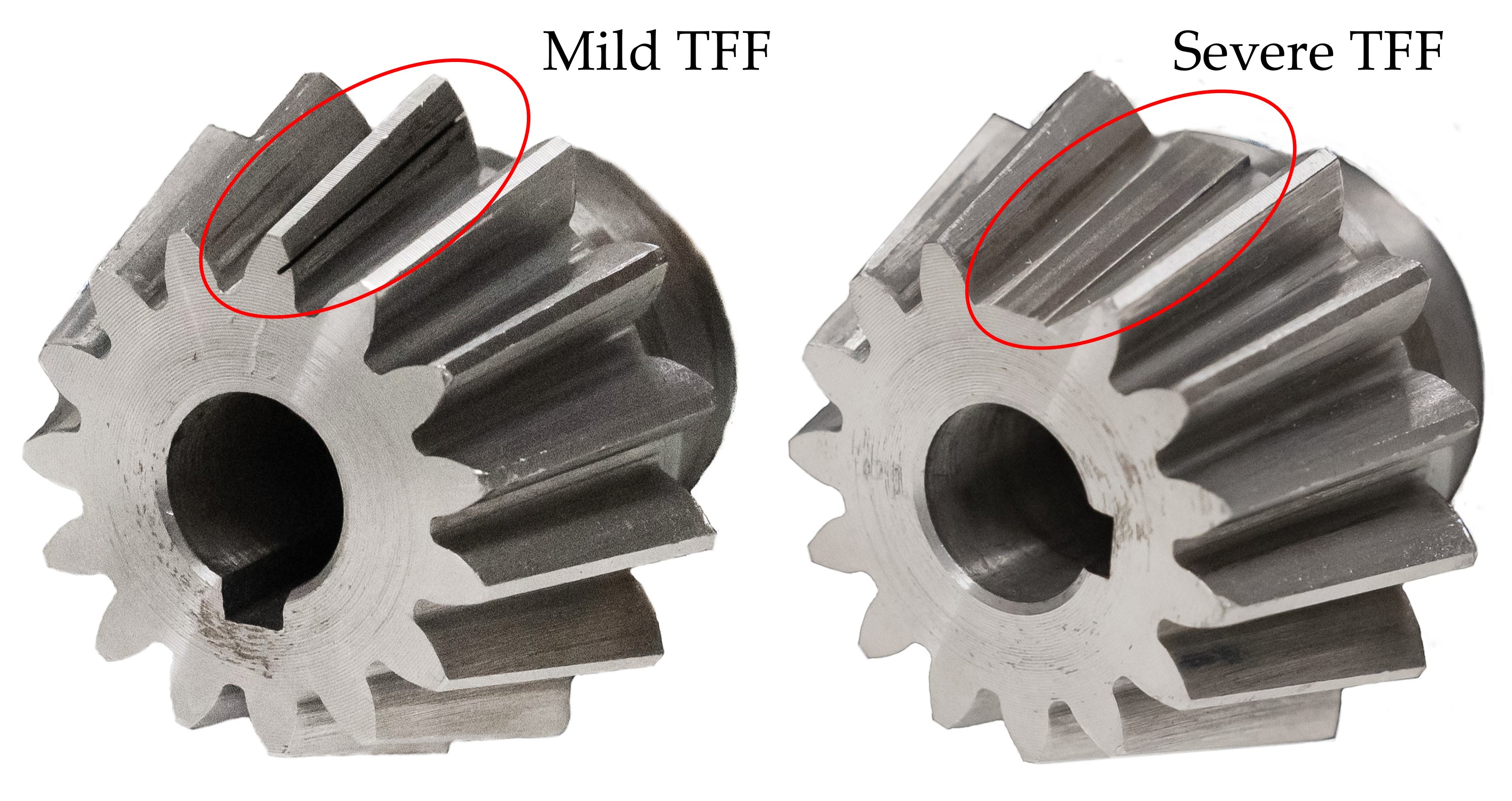}
            \vspace{-4mm}
            \caption{Faulty pinions with a mild and severe TFF.}
            \label{fig:pn_tff}
        \end{figure}

    The lower gearbox was specifically designed for gear defect experimentation, featuring a straight-tooth bevel gear with a driving pinion of 15 teeth and a driven gear wheel of 60 teeth, both of module three. In addition to healthy conditions, several fault cases have been measured, of which mild and severe TFF are selected for this study. The mild TFF is a 1 mm wide void extending across the full width of one pinion tooth, and the severe TFF is demonstrated by a missing tooth (Figure \ref{fig:pn_tff}). TFF has emerged as one of the main failure mechanisms in modern gear design, especially in high-speed, high-power gear systems \cite{rommel2025}. 

    The test rig sensors used in this study were two strain gauge-based torque transducers and five incremental rotary encoders, enabling the precise monitoring of torsional behaviour. The system synchronously samples all sensors at a rate of 3012 Hz. The first torque transducer is installed on the drive shaft before the upper bevel gear and the second on the propeller shaft after the lower bevel gear (Figure \ref{fig:aalto_gear}). Two of the rotary encoders are located on the drive shaft, two on the middle shaft, and one on the propeller shaft.
    
    \subsection*{Measurements}
    The measurement data used in this study focuses on three health conditions: healthy, mild pinion TFF, and severe pinion TFF \cite{dahl2024}. The drive motor operated at six rotational speeds ranging from 250 to 1500 RPM in 250 RPM increments. These speeds correspond to propeller axis speeds between 21 RPM and 125 RPM. For each motor speed, the load motor was applied at three torque levels: 1\%, 6\%, and 11\% of its nominal torque, corresponding to 0.12 Nm, 0.71 Nm, and 1.31 Nm, respectively.

    Each pinion was paired with a dedicated gear wheel to form three consistent gear pairs. Baseline tests were first conducted with healthy gear pairs. After introducing gear faults to the pinions, the same gear pairs were reassembled, and measurements were repeated in their faulty states. The lowest characteristic torsional frequency of the experimental system is 24 Hz.

\section{Simulation model} \label{sec:sect_sim}
The thruster drivetrain system was modelled by using the MBS approach to demonstrate the simulation of torsional phenomena, particularly torsional vibrations. To observe the related effects from the lower bevel gearbox in faulty conditions, we implemented the gear meshing with the PCM, available in the Simpack software. The complete simulation system represents three rotating drivelines interconnected by two bevel gearboxes. At this stage, the PCM was only used to model contacts in the lower gearbox, as it is the one containing the faults. Simplifications and limitations regarding the simulation model are listed as follows:

\begin{itemize}
    \item The drive and load are modelled as external rotation through speed and torque load inputs with their inertias, excluding the mechanics and electromagnetics of the motors.
    \item The upper gearbox is modelled with a simplified gear transmission element that does not take gearwheel geometries into account.
    \item The lubrication in the lower gearbox is modelled by using Coulomb’s friction, with a lower friction coefficient.
    \item To analyse torsional vibrations, only the longest and thinnest rotating shafts are modelled as flexible (green shafts in Figure \ref{fig:simpack_thruster}). while other components are treated as rigid bodies. The other shafts are sufficiently stiff that their dynamic effects become significant only at much higher frequencies, which are not relevant for gear mesh analysis.
\end{itemize}

    \subsection{Geometries}
\begin{figure}[H]
        \centering
        \includegraphics[width=0.73\linewidth, trim={0.1cm 0.1cm 0.1cm 0.1cm},clip]{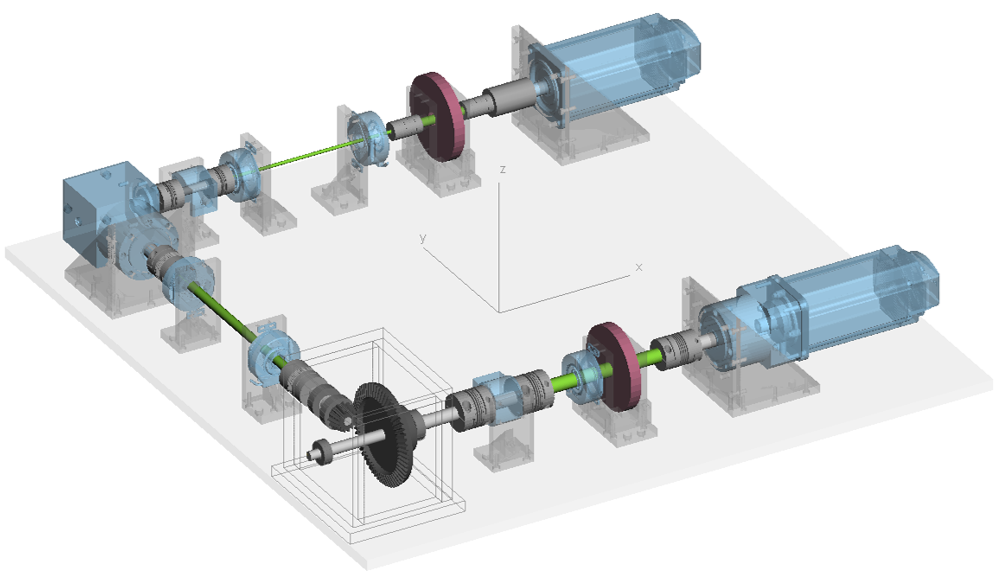}
        \vspace{-3mm}
        \caption{The full simulation model of the downscaled thruster test rig. The rotating components are coloured as opaque colours, of which the flexible shafts are green. Non-rotating parts are semi-transparent.}
        \label{fig:simpack_thruster}
    \end{figure}

    Component geometries were decided based on their functional role and influence on system dynamics. The simplification of less critical components has a direct impact on numerical performance. Figure \ref{fig:simpack_thruster} illustrates the full simulation system of the downscaled thruster drivetrain. Rotating components, namely shafts, flywheels, couplings, and lower gearbox gearwheels, are shown as opaque bodies. These were modelled by using geometrical dimensions and material properties to ensure the correct representation of mass and inertia. The coordinate system was defined so that the driven and propeller shafts rotate about the X-axis, and the middle shaft about the Y-axis. 
    
\begin{figure}[H]
        \centering
        \includegraphics[width=0.5\linewidth, trim={0.4cm 0.3cm 0.4cm 0.4cm},clip]{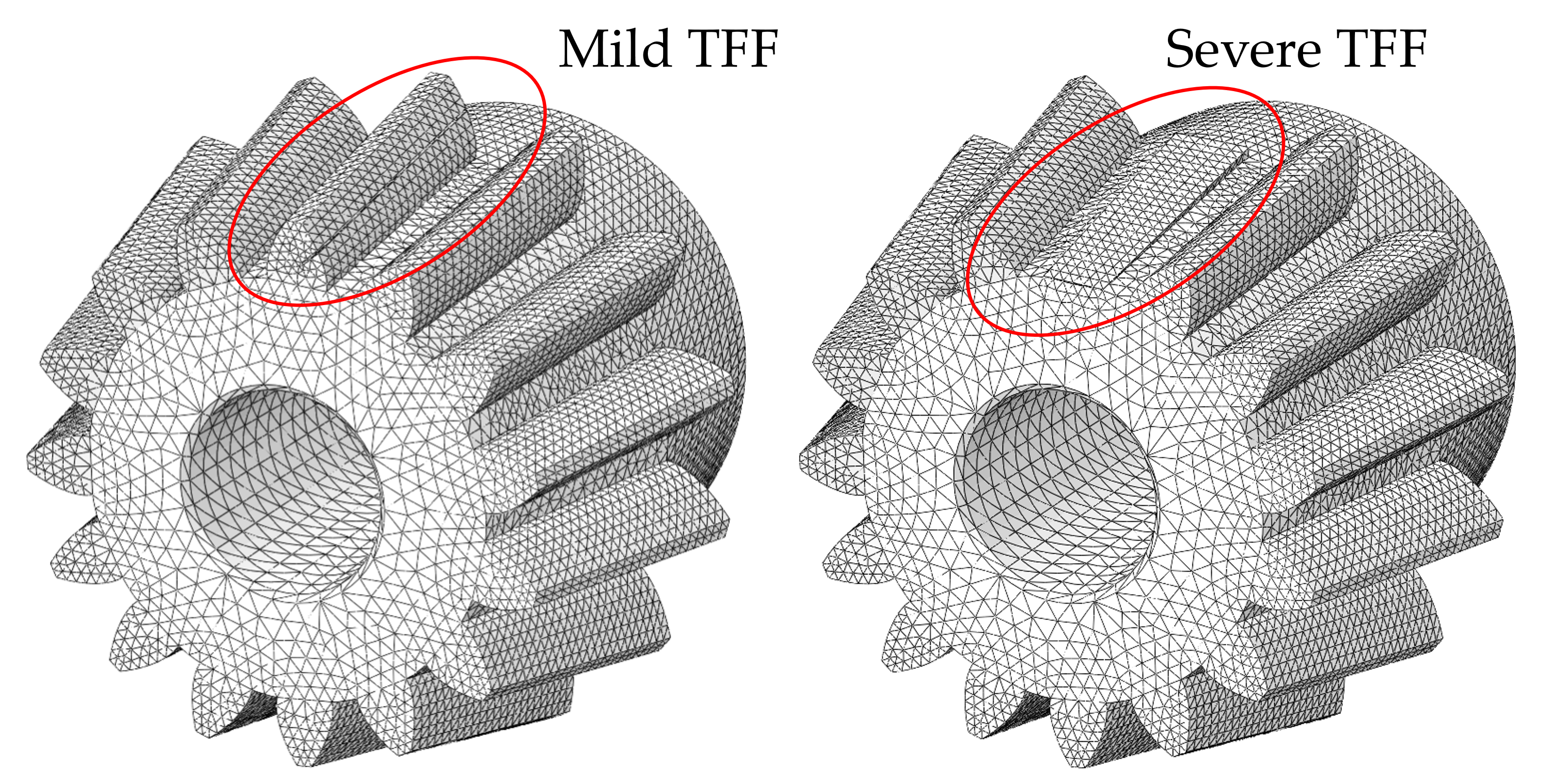}
        \vspace{-3mm}
        \caption{Faulty pinion geometries in the lower bevel gearbox used for simulating the mild and severe TFF. Polygon size on the contacting surfaces, namely on teeth, is 0.5 mm.}
        \label{fig:pinion_meshes}
    \end{figure}

    Shafts and flywheels were defined as simple cylinder shapes, while couplings were imported as simplified CAD geometries, supplied by the manufacturer. The lower gearbox gearwheels were represented as polygonal surface models imported in wavefront objects (.obj), a format that encodes 3D geometry through polygonal meshes. Detailed gear dimensions were derived by combining knowledge of standard gear design practices, confirmed by measurements from the actual gearwheels. A polygon size of 0.5 mm was selected for contact surfaces (Figure \ref{fig:pinion_meshes}). 

    Four shafts (green in Figure \ref{fig:simpack_thruster}) are modelled by using SIMBEAM bodies, three-dimensional flexible beam structures within Simpack \cite{simpackmanual2024_simbeam}. Flexible bodies ensured that the lowest torsional characteristic frequencies were captured accurately.

    Geometries of non-rotating parts (semi-transparent in Figure \ref{fig:simpack_thruster}) were primarily included for visual reference. The upper gearbox and motors were represented by their respective rotor inertias to account for mass effects, with the load motor additionally including the inertia of the planetary gearbox. Similarly, sensors and bearings were implemented as special force elements with visual-only geometries, as their masses do not affect system dynamics.

    \subsection{Connections and force elements}
    Non-rotating parts, including gearbox housings, motors, and sensors, were individually mounted to a rigid platform at equal shaft heights throughout the drivetrain. Rotating shafts were connected via flexible couplings, represented as bushing elements with defined torsional, axial, and radial stiffness and damping. These parameters were either calculated by using finite element method (FEM) or provided by manufacturers. Moreover, flywheels and lower gearbox gearwheels were rigidly fixed to their respective shafts.
    
    The system considered eight bearings: Four ball bearings supporting the flywheel shafts, and four tapered rolling bearings supporting the gearwheel shafts in the lower gearbox. As bearing faults are beyond the scope of this study, their dynamics were simulated by using simplified rolling bearing elements that mimic typical healthy bearing dynamics, including torque losses. The upper gearbox, assumed to be healthy, was modelled by using a simplified gear transmission element. This element incorporated rotational inertias for the drive and middle shafts and a gear transmission ratio of 3:1. For the lower gearbox, with a transmission ratio of 4:1, gearwheel contacts were modelled by using PCM.

    To provide a foundation for healthy gear contact, we validated the surface geometry-based PCM (FE199) with Gear Pair module (FE225). The Gear Pair is a Hertzian-based element for modelling gears through a detailed contact calculation between two meshing gear wheels with involute geometry. This Simpack element is specifically designed for analytical contact calculation of gear meshing forces based on tooth penetration, incorporating Hertzian contact stiffness through models such as DIN 3990 or Weber/Banaschek \cite{simpackmanual2024_fe225}.

    The driving torque applied to the motor shaft was controlled by a proportional-integral (PI) controller in a feedback loop, maintaining shaft speed at constant velocities. The load at the load motor shaft was modelled as velocity-dependent viscous torque, accounting for the final transmission ratios. Each velocity-load-fault combination was simulated for 20 seconds with a sampling rate of 3 kHz. Finally, sensor elements recorded data corresponding to experimental measurements, capturing rotational angles and torsional vibrations analogous to encoder and torquemeter outputs, respectively.
    
    \subsection{System validation}
    The thruster drivetrain dynamics were carefully validated step by step for the healthy system by using valid computational methods as well as the experimental measurements. In the initial development stage, both gearboxes were modelled by using simplified transmission elements with ideal transmission ratios without detailed gearwheel geometry. This simplification aimed to validate the fundamental torsional dynamics at the system level before introducing further complexity. 

    Since simulation models assume ideal torque transmission, establishing accurate average torque levels throughout the driveline required incorporating realistic power losses, as defined by Equations~\eqref{eq:power} and~\eqref{eq:power_loss}. Although isolating individual loss mechanisms is challenging, total power losses can be quantified from experimental measurements. Some of the simulated drivetrain components, such as bearing bushings and gear elements, already account for comparable physical losses. The remaining losses were approximated by using speed-dependent torsional bushings that represent viscous damping. These bushing elements were added at the planetary gearbox and the upper gearbox and were calibrated to reproduce the average torque reductions observed based on the measurements.

    Table~\ref{tab:torque_losses} lists the average measured torque and constant rotational speed at the drive shaft (DSH) and propeller shaft (PSH), along with the calculated power losses. The total power loss $P_{tot}$, between the torquemeter locations, and the losses attributed to the planetary gearbox $P_{pg}$, were implemented in the simulation model by adjusting the viscous damping parameters accordingly. In this way, we were able to produce simulated torque levels, comparable to measured data at each operating point.
    
    \begin{table}[width=0.8\linewidth,cols=9,pos=h!]
    \vspace{-5mm}
    \caption{Overview of measured average torque, speed, and power values at different load levels.}
    \label{tab:torque_losses}
    \centering
    \begin{tabular*}{\tblwidth}{@{} CCCCCCCCC@{} }
        \toprule
        $\bm{T}_{\text{load}}$ & $\bm{T}_{\text{DSH}}$ & $\bm{T}_{\text{PSH}}$ & $\bm{\omega}_{\text{DSH}}$ & $\bm{\omega}_{\text{PSH}}$ & $\bm{P}_{\text{DSH}}$ & $\bm{P}_{\text{PSH}}$ & $\bm{P}_{\text{pg}}$ & $\bm{P}_{\text{tot}}$ \\
         \midrule
        1.31 & 1.85 & 14.75 & 1502.98 & 125.25 & 291.43 & 193.42 & 137.46 & 98.00 \\
        1.31 & 1.87 & 14.84 & 1252.48 & 104.37 & 245.39 & 162.20 & 114.54 & 83.18 \\
        1.31 & 1.80 & 14.58 & 1001.98 & 83.50  & 188.65 & 127.51 & 91.64  & 61.14 \\
        1.31 & 1.74 & 14.45 & 751.49  & 62.62  & 137.18 & 94.77  & 68.72  & 42.42 \\
        1.31 & 1.63 & 13.93 & 500.99  & 41.75  & 85.56  & 60.92  & 45.82  & 24.63 \\
        1.31 & 1.51 & 13.22 & 250.50  & 20.87  & 39.54  & 28.88  & 22.90  & 10.66 \\ 
        0.71 & 1.44 & 10.02 & 1502.98 & 125.25 & 226.75 & 131.45 & 74.50  & 95.30 \\
        0.71 & 1.41 & 9.81  & 1252.48 & 104.37 & 185.48 & 107.26 & 62.08  & 78.22 \\
        0.71 & 1.36 & 9.72  & 1001.98 & 83.50  & 143.01 & 84.97  & 49.67  & 58.05 \\
        0.71 & 1.30 & 9.48  & 751.49  & 62.62  & 101.98 & 62.15  & 37.25  & 39.83 \\
        0.71 & 1.20 & 8.97  & 500.99  & 41.75  & 62.76  & 39.20  & 24.83  & 23.57 \\
        0.71 & 1.08 & 8.33  & 250.50  & 20.87  & 28.23  & 18.20  & 12.41  & 10.03 \\ 
        0.12 & 1.03 & 5.24  & 1502.98 & 125.25 & 161.85 & 68.67  & 12.59  & 93.18 \\
        0.12 & 0.98 & 4.97  & 1252.48 & 104.37 & 128.43 & 54.30  & 10.49  & 74.13 \\
        0.12 & 0.94 & 4.88  & 1001.98 & 83.50  & 98.37  & 42.68  & 8.39   & 55.69 \\
        0.12 & 0.86 & 4.57  & 751.49  & 62.62  & 67.64  & 29.95  & 6.30   & 37.69 \\
        0.12 & 0.77 & 4.09  & 500.99  & 41.75  & 40.28  & 17.86  & 4.20   & 22.42 \\
        0.12 & 0.65 & 3.44  & 250.50  & 20.87  & 16.93  & 7.52   & 2.10   & 9.41  \\ 
        \bottomrule
    \end{tabular*}
\end{table}

    Subsequently, we performed eigenvalue analysis on the full system to verify that the lowest natural frequencies and corresponding mode shapes, primarily influenced by the thinnest shafts and coupling connections, were consistent with previous findings. Our simulation model yielded the lowest torsional frequency of approximately 25 Hz, closely matching the one-dimensional calculations reported in \cite{laine2025a}.

    Once losses and system frequencies were confirmed, we replaced the simple transmission element in the lower bevel gearbox with actual gearwheel geometries (FE199 \& FE225), while carefully maintaining the basic dynamic behaviour of the system. In parallel with comparing simulation results to measurements, the healthy PCM parameters (FE199) were iterated to correspond to the simpler Gear Pair module (FE225). For example, we checked that tooth contacts, either one or two teeth in contact, and the maximum contact pressure, were comparable. This way the sufficient polygon size for the gearwheels was also determined. 

\begin{figure}[H]
        \centering
        \includegraphics[width=0.73\linewidth, trim={0.2cm 0.3cm 0.3cm 0.2cm},clip]{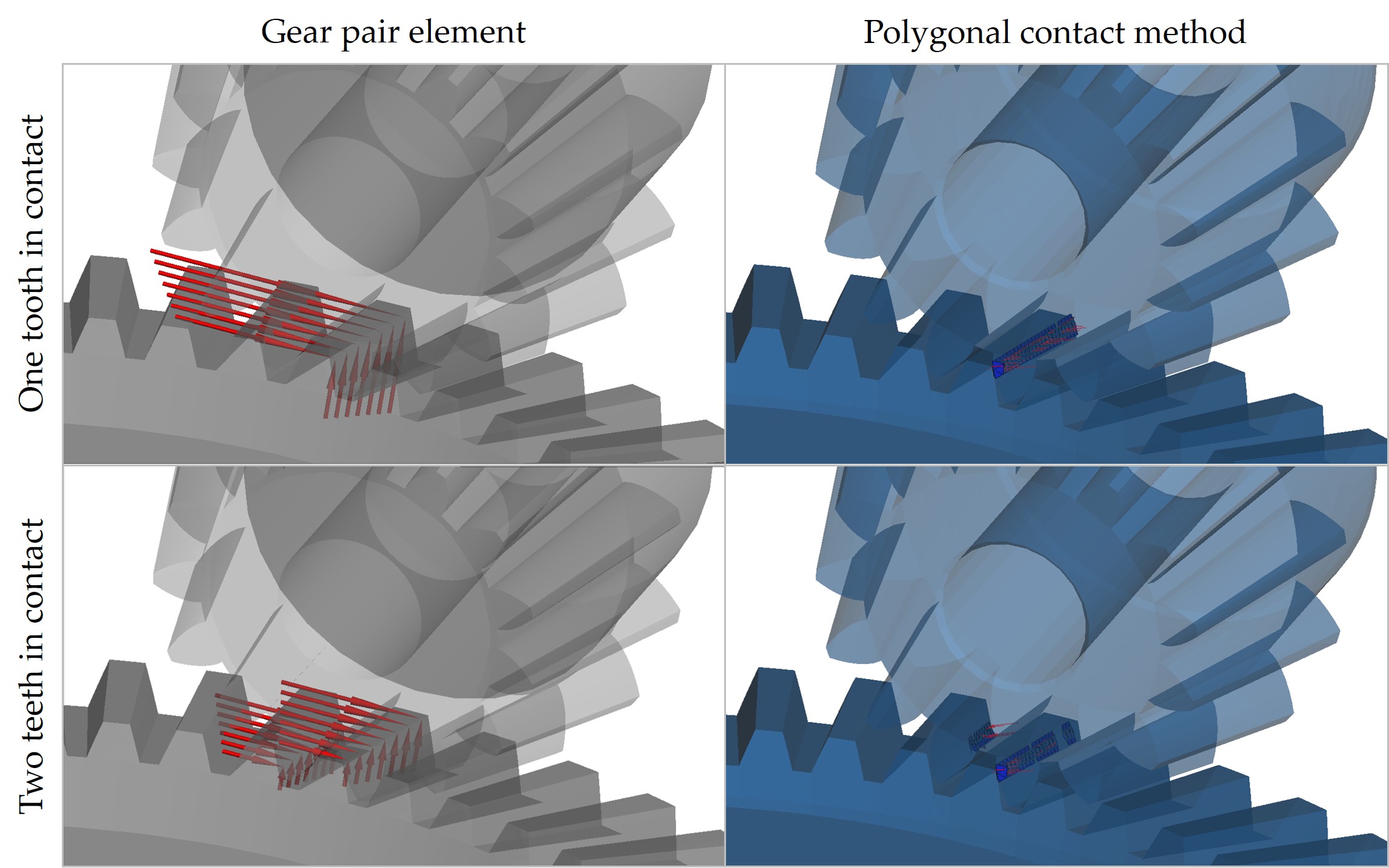}
        \vspace{-3mm}
        \caption{Three-dimensional bevel gear meshing illustrated with the Gear Pair (FE225) and PCM (FE199) elements, shown during both single and double tooth contact phases.} \label{fig:simpack_contacts}
        \vspace{-2mm}
    \end{figure}

\begin{figure}[H]
        \centering
        \includegraphics[width=0.85\linewidth,clip]{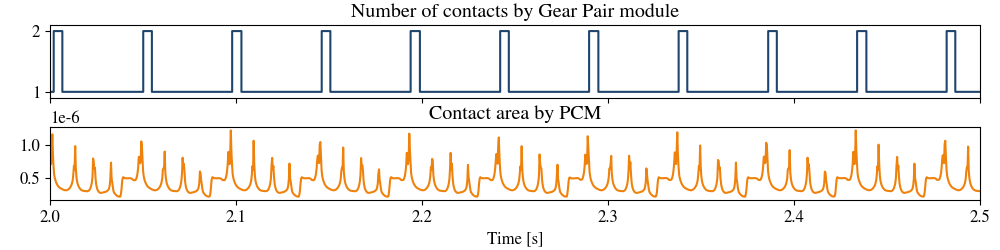}
        \vspace{-3mm}
        \caption{Number of teeth in contact from the Gear Pair (FE225) compared to the geometry-based contact area calculated by PCM (FE199).} \label{fig:contact_graphs}
    \end{figure}

    Figure \ref{fig:simpack_contacts} visualises how the Gear Pair module distributes tooth contact axially over a user-defined number of slices, whereas PCM defines the contact area directly from the CAD geometry. Similarly, Figure \ref{fig:contact_graphs} presents the natural fluctuation between single and double tooth contact captured by the Gear Pair. The pattern is also observed in the PCM model, where the geometry-based contact area increases during double tooth engagement.
    
\begin{figure}[H]
        \centering
        \includegraphics[width=0.85\linewidth,clip]{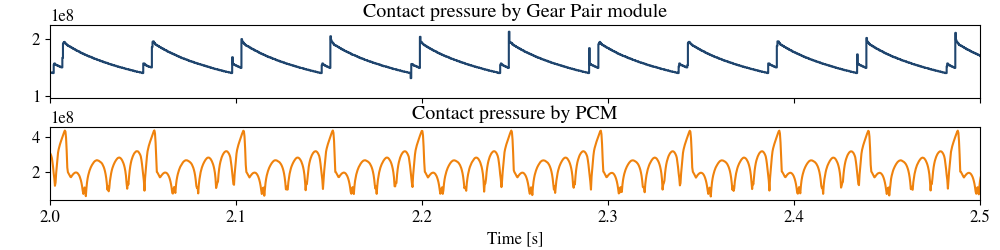}
        \vspace{-3mm}
        \caption{Comparison of contact pressure between the Gear Pair module (FE225) and the PCM (FE199).} \label{fig:pressure_graphs}
    \end{figure}

    The Gear Pair computes Hertzian-based contact pressure that shows periodic single and double tooth engagement illustrated in Figure \ref{fig:pressure_graphs}. The PCM model reflects similar fluctuations in the maximum contact pressure, capturing additional polygonal geometry-related local phenomena. The final PCM contact properties are listed in Table \ref{tab:pcm_specs}. After the PCM was ready for healthy simulations, faulty cases were simulated just by changing the pinion geometry.

    \begin{table}[width=0.6\linewidth,cols=3,pos=h!]
    \caption{PCM properties applied in the lower bevel gear meshing.}
    \label{tab:pcm_specs}
    \centering
    \begin{tabular*}{\tblwidth}{@{} CCC@{} }
        \toprule
        \textbf{Denotation} & \textbf{Parameter}                        & \textbf{Value} \\ 
        \midrule
        $\nu_{1,2}$         &   Poisson's ratio                         &   0.3 \\
        $E_{1,2}$           &   Young's modulus                         &   206 GPa \\
        $b_{1,2}$           &   Elastic layer depth                     &   0.5 mm \\ 
        $c_c$               &   Compression-related damping             &   4e4  MNs/m\textsuperscript{3} \\  
        $c_e$               &   Expansion-related damping               &   3e3  MNs/m\textsuperscript{3} \\
        $t_d$               &   Transformation depth for damping        &   5 \textmu m  \\
        $\mu$               &   Friction coefficient                    &   0.002 \\
        $v_f$               &   Regularised velocity for friction       &   0.001 m/s \\
        $u_{max}$           &   Maximum penetration                     &   1 mm \\ 
        \bottomrule
    \end{tabular*}
\end{table}

\section{Results} \label{sec:sect_results}
We developed a system-level MBS model of the full thruster drivetrain, comprising of three shafts: drive shaft (DSH), middle shaft and propeller shaft (PSH). Contacts between the lower gearbox gearwheels were implemented with the PCM. Our results present measured and simulated torsional vibration signals from the drive shaft and propeller shaft in both time and frequency domains. 

Altogether, 54 simulations were performed, covering six rotational speeds, three torque loads, and three health conditions. We pre-processed all data by applying Hanning's windowing and low-pass filtering with a cut-off of five times the GMF. Given the large volume of data generated for CM purposes, only representative cases to address the main results are presented. The time-domain analysis focuses on two key speeds, 500 RPM and 1500 RPM, at a constant load torque of 1.31 Nm, comparing healthy, mild TFF, and severe TFF conditions. For the frequency-domain analysis, we initially examine the full spectra for these same representative cases. Additionally, we analyse results for three speeds (250, 1000, and 1500 RPM), across three health conditions and two load levels, with particular attention to theoretical fault frequencies.

    \subsection{Time domain signals}
\begin{figure}[H]
        \centering
        \includegraphics[width=0.49\linewidth, trim={0cm 0cm 0cm 1.4cm},clip]{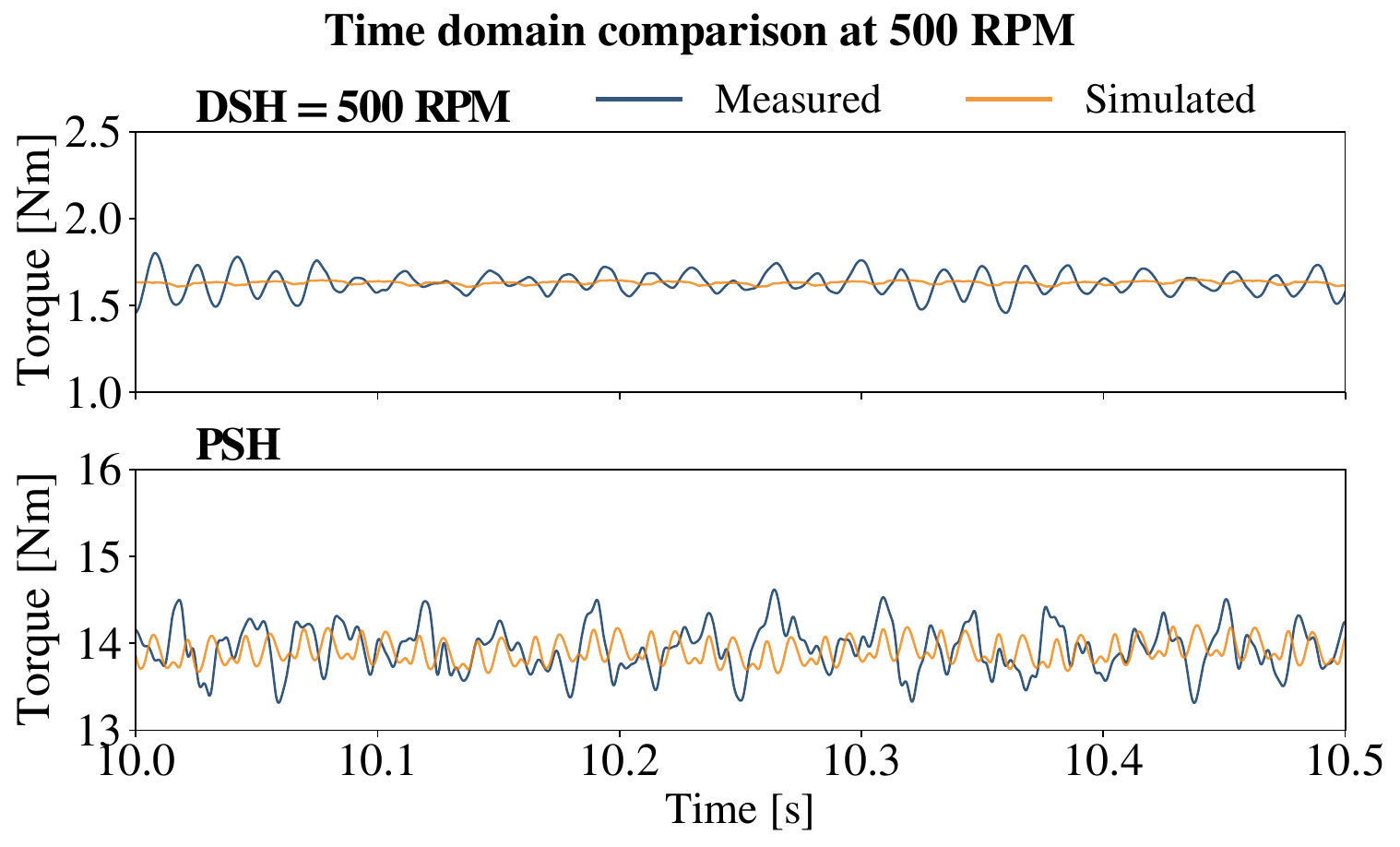}
        \includegraphics[width=0.49\linewidth, trim={0cm 0cm 0cm 1.4cm},clip]{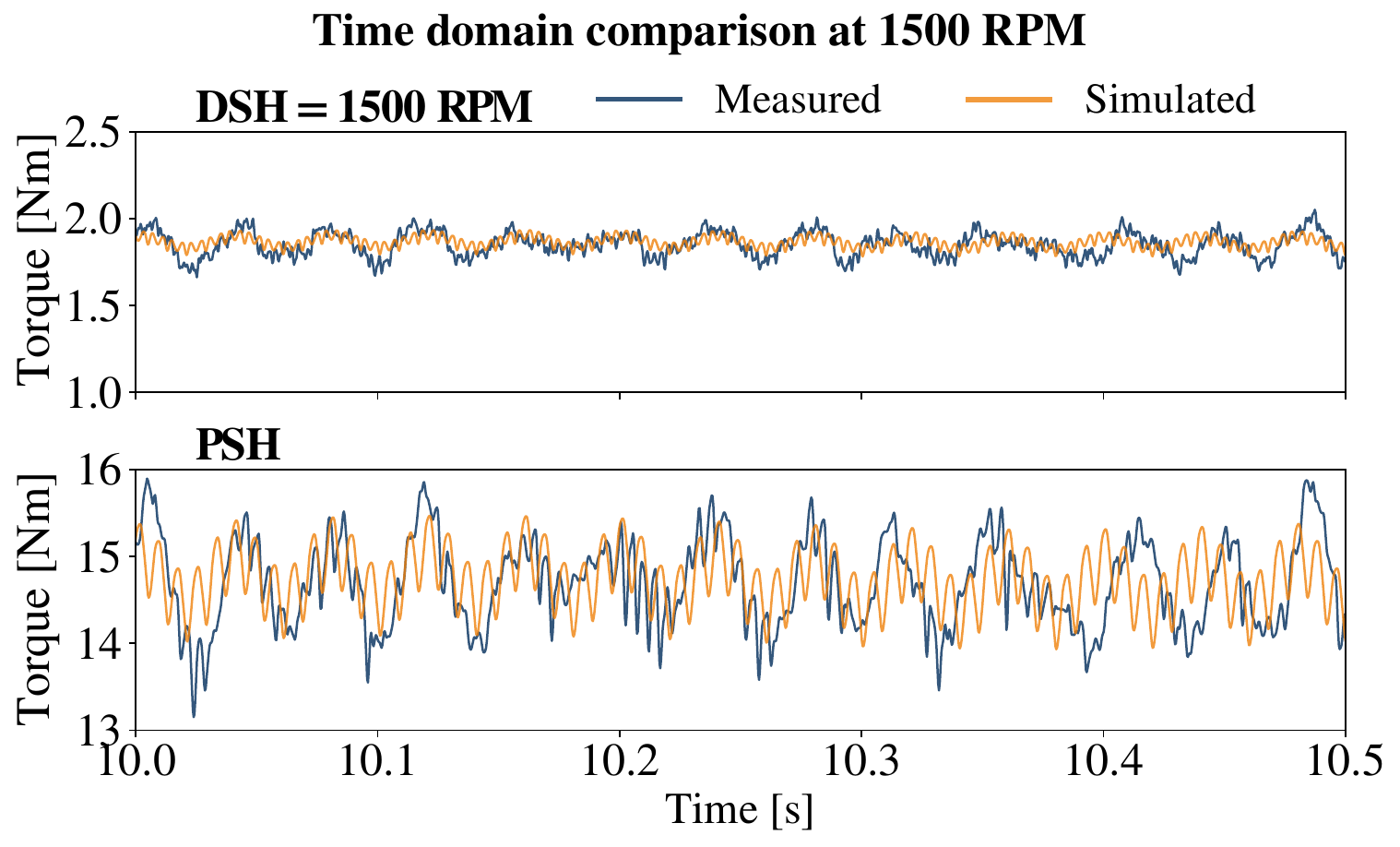}
        \vspace{-2mm}
        \caption{\label{fig:td_healthy} PSH and DSH torque waveforms with a healthy pinion at 500 RPM and 1500 RPM drive speed. The torque load is 1.31 Nm.}
    \end{figure}

\begin{figure}[H]
        \centering
        \includegraphics[width=0.49\linewidth, trim={0cm 0cm 0cm 1.4cm},clip]{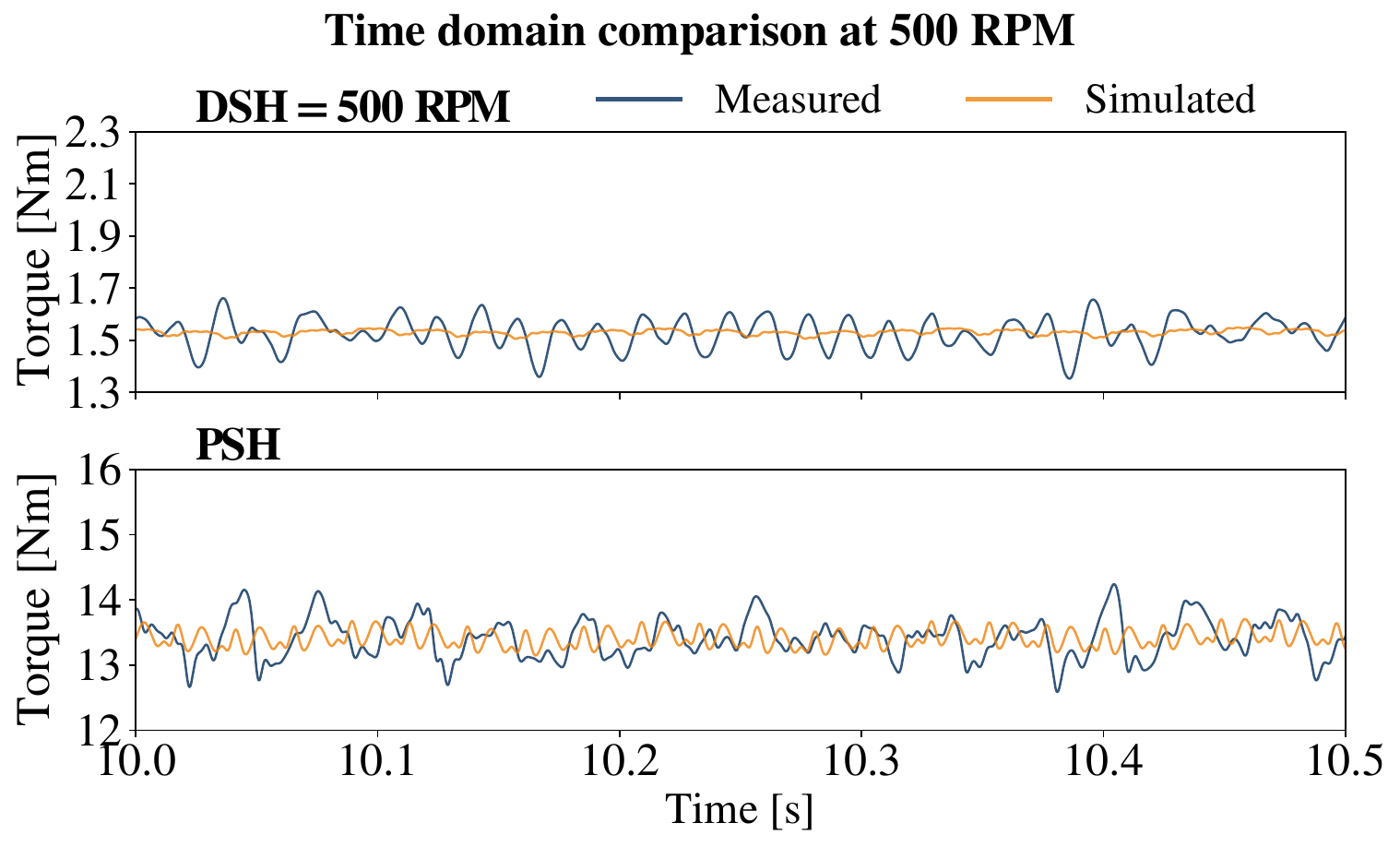}
        \includegraphics[width=0.49\linewidth, trim={0cm 0cm 0cm 1.4cm},clip]{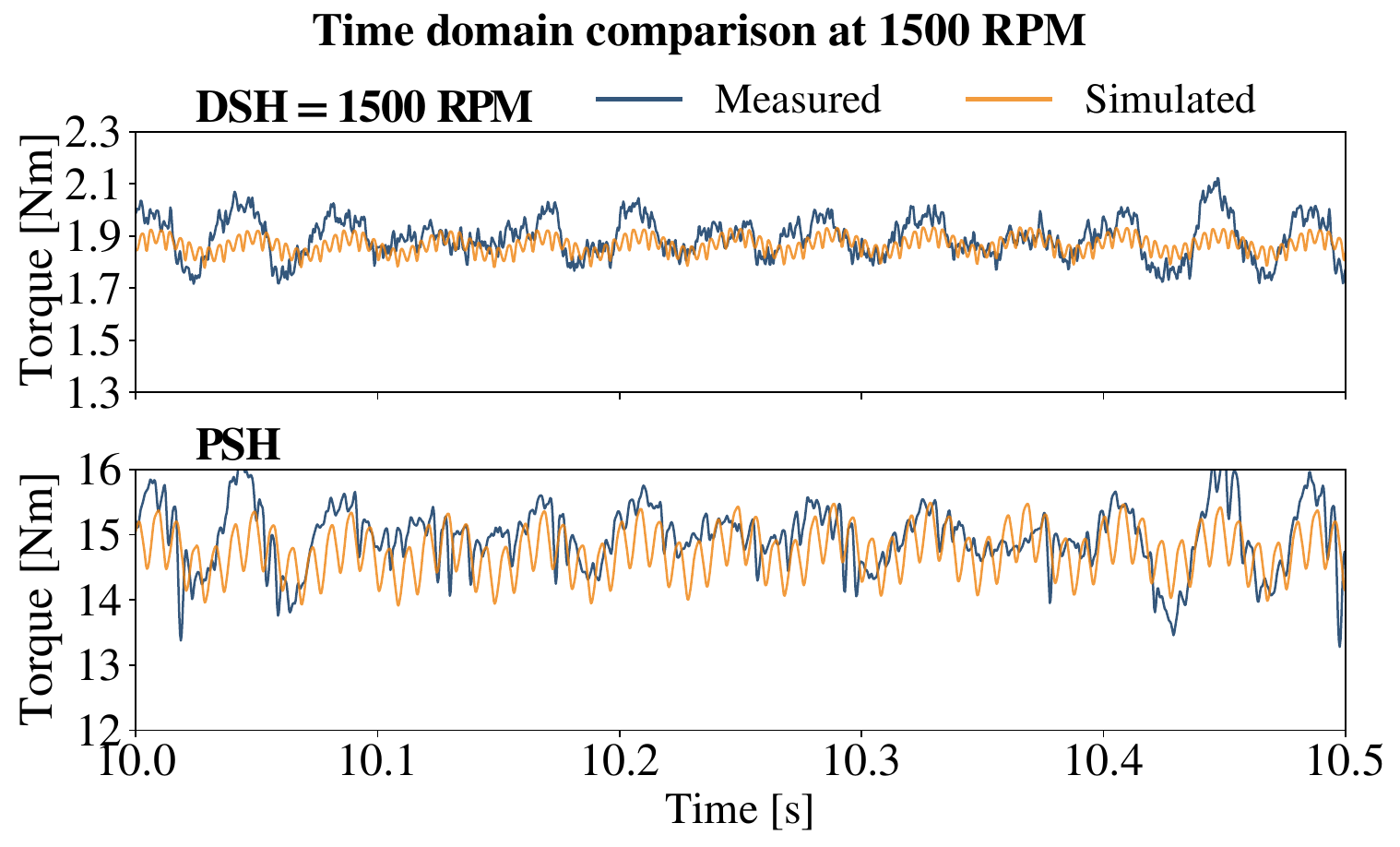}
        \vspace{-2mm}
        \caption{\label{fig:td_mild_tff} PSH and DSH torque waveforms with a mild pinion TFF at 500 RPM and 1500 RPM drive speed. The torque load is 1.31 Nm.}
    \end{figure}

\begin{figure}[H]
        \centering
        \includegraphics[width=0.49\linewidth, trim={0cm 0cm 0cm 1.4cm},clip]{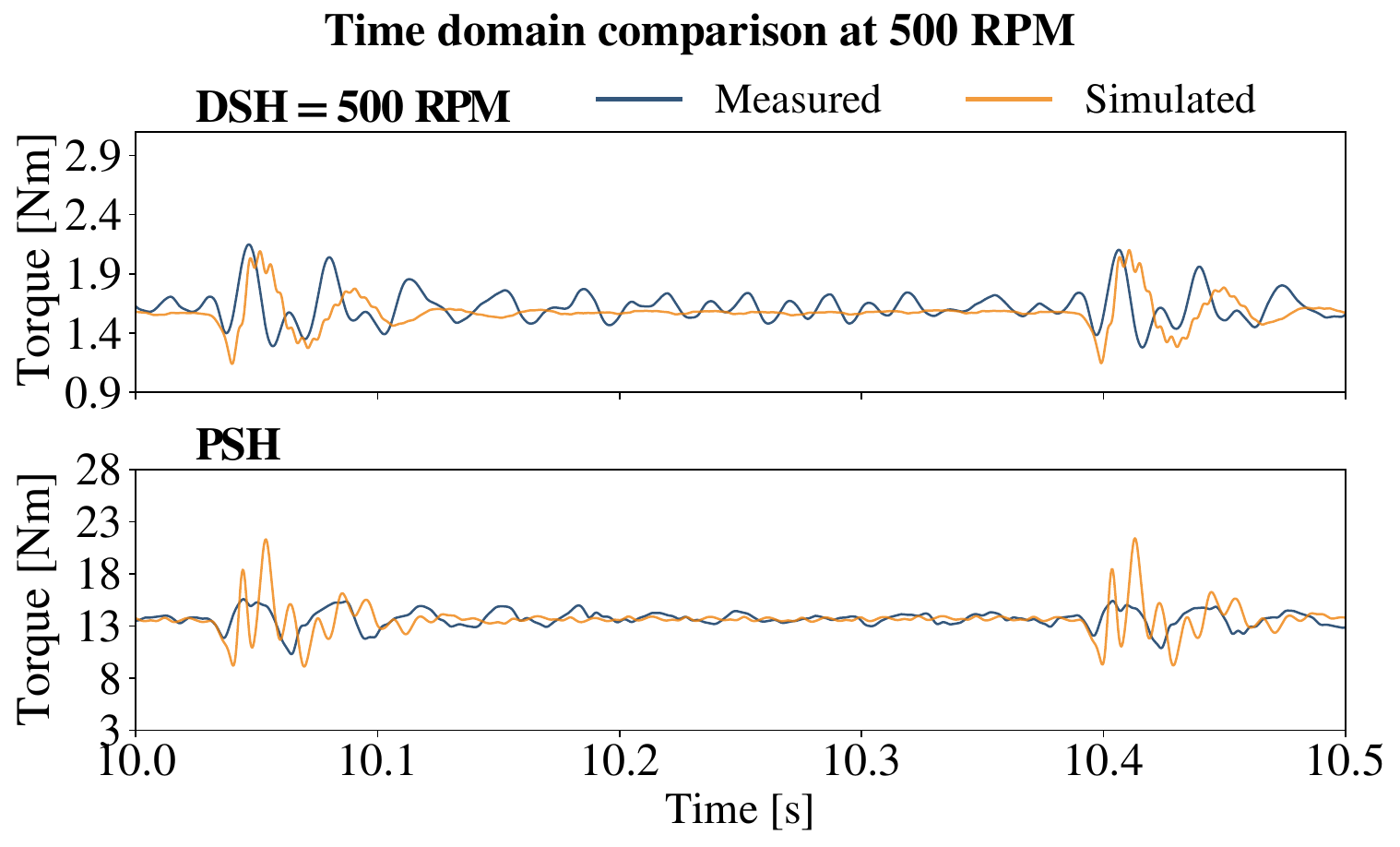}
        \includegraphics[width=0.49\linewidth, trim={0cm 0cm 0cm 1.4cm},clip]{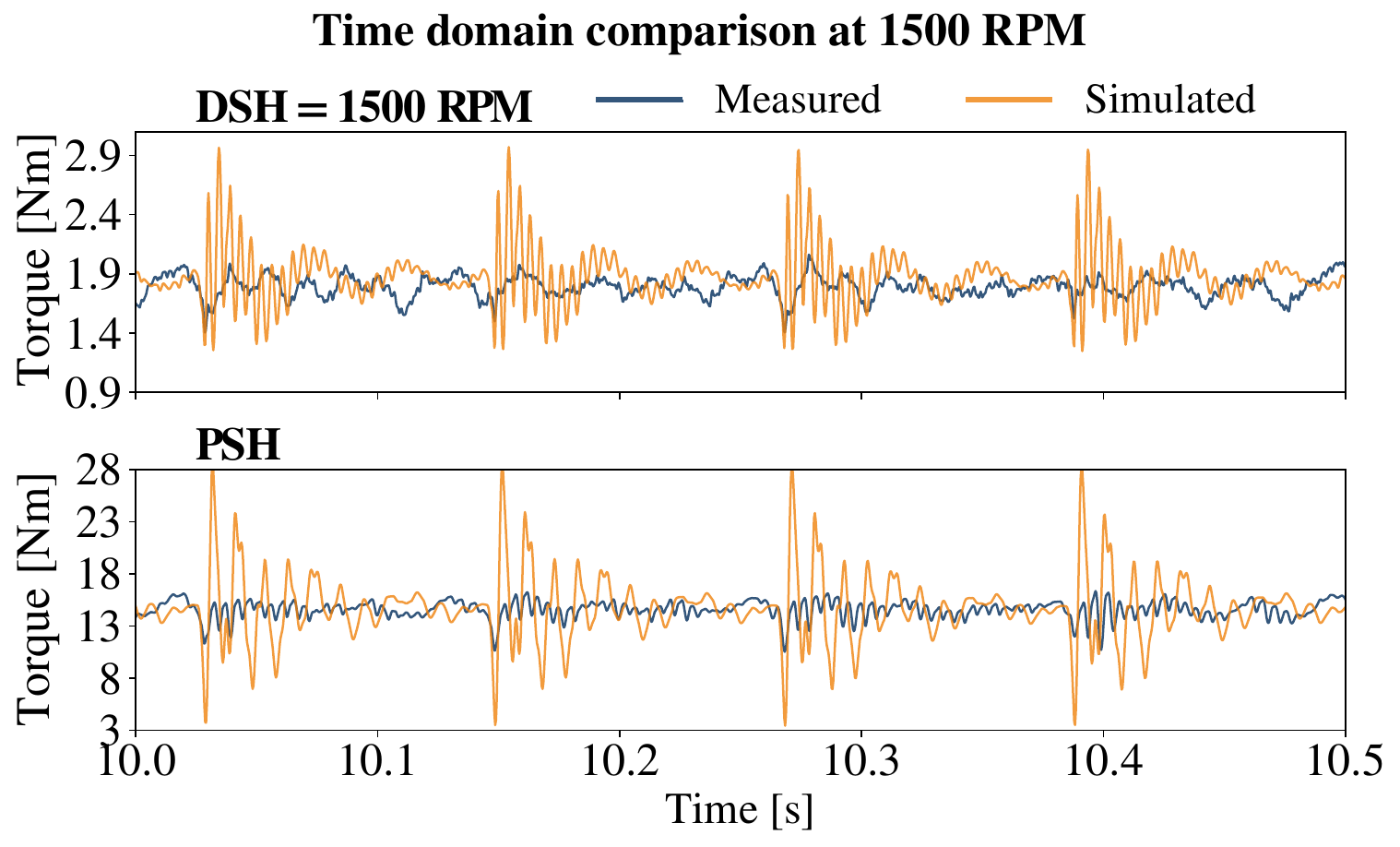}
        \vspace{-2mm}
        \caption{\label{fig:td_severe_tff} PSH and DSH torque waveforms with a severe pinion TFF at 500 RPM and 1500 RPM drive speed. The torque load is 1.31 Nm.}
    \end{figure}

    In the time domain, we analysed torque waveforms, measured and simulated at the drive shaft and propeller shaft. The resulting waveforms for both representative speeds are shown in Figures \ref{fig:td_healthy}–\ref{fig:td_severe_tff}. Generally, simulated signals exhibit more regular waveforms with lower torque variations than their measured counterparts, which is expected due to the idealised conditions inherent to simulations. The average simulated torque levels for both drive and propeller shafts closely match measured torque levels, and increased rotational speed results in slightly higher average torque and greater torque variations, confirming an accurate representation of gear transmission and system losses in the simulation model. However, simulated drive shaft torque variations are notably lower than measured values, particularly at higher speeds where measured amplitudes significantly increase. This discrepancy suggests potential eccentricities or irregularities present during measurements, which the simulation cannot fully replicate. Nevertheless, simulated drive shaft torque does indicate a relative increase towards higher speeds.

    In the healthy condition (Figure \ref{fig:td_healthy}), measured and simulated signals display consistent periodic oscillations related to gear meshing and shaft rotation. Propeller shaft torque waveforms are more comparable, although measured signals exhibit slightly damped, less frequent high-frequency oscillations. This likely results from real-world damping and structural properties not fully captured by the simulation. Measured data also show greater peak-to-peak amplitudes and irregularities, indicative of minor unmodelled effects such as measurement noise and manufacturing deviations. While simulations accurately predict primary torsional behaviour, they seem to underestimate higher-frequency disturbances present in measurements.

    With mild TFF (Figure \ref{fig:td_mild_tff}), both measured and simulated signals retain recognisable periodicity, though amplitude modulations become slightly more evident compared to the healthy state, especially in measured data. Although simulations reflect increased modulation trends, they fail to fully capture the magnitude and irregularity seen in measured signals. A higher rotational speed amplifies amplitude modulations and transient disturbances, but these remain subtle and less pronounced in simulations when compared to measurements. Overall, the effect of the mild fault on the torque signal is still barely visible.
    
    Under severe TFF (Figure \ref{fig:td_severe_tff}), measured and simulated torque signals deviate significantly from the healthy periodic behaviour, exhibiting strong transient events and substantial amplitude spikes. Increasing the rotational speed amplifies these transient events, prolonging the time required for torque signals to stabilise. Simulations closely replicate the timing and occurrence of these transients but consistently produce higher amplitude spikes and faster stabilisation when compared to measured data. This behaviour implies an overestimation of stiffness or insufficient damping representation in the simulation model, highlighting simplified complexities, such as lubrication effects, system compliance, and other dynamic factors influencing at higher speeds and with more severe faults. In contrast, measured signals display more moderate peaks and distributed transient impacts.
    
    \subsection{Frequency domain analysis}
\begin{figure}[H]
        \centering
        \includegraphics[width=\linewidth, trim={0cm 0cm 0cm 1.4cm},clip]{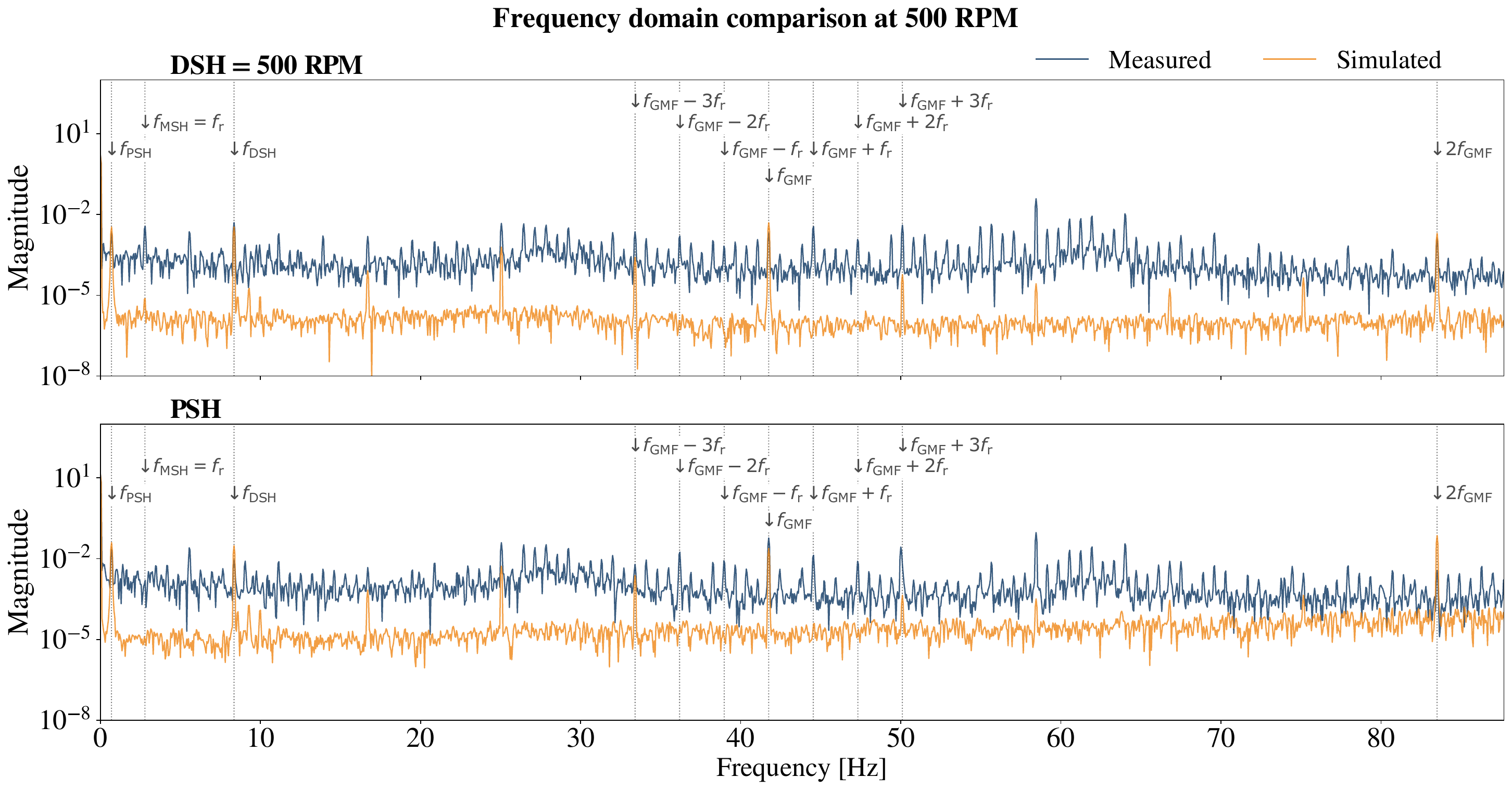} \vspace{-7mm}
        \caption{\label{fig:fft_healthy_500rpm} DSH and PSH torque spectra with a healthy pinion at 500 rpm drive shaft speed and 1.31~Nm torque load.}
    \end{figure}

\begin{figure}[H]
        \centering
        \includegraphics[width=\linewidth, trim={0cm 0cm 0cm 1.4cm},clip]{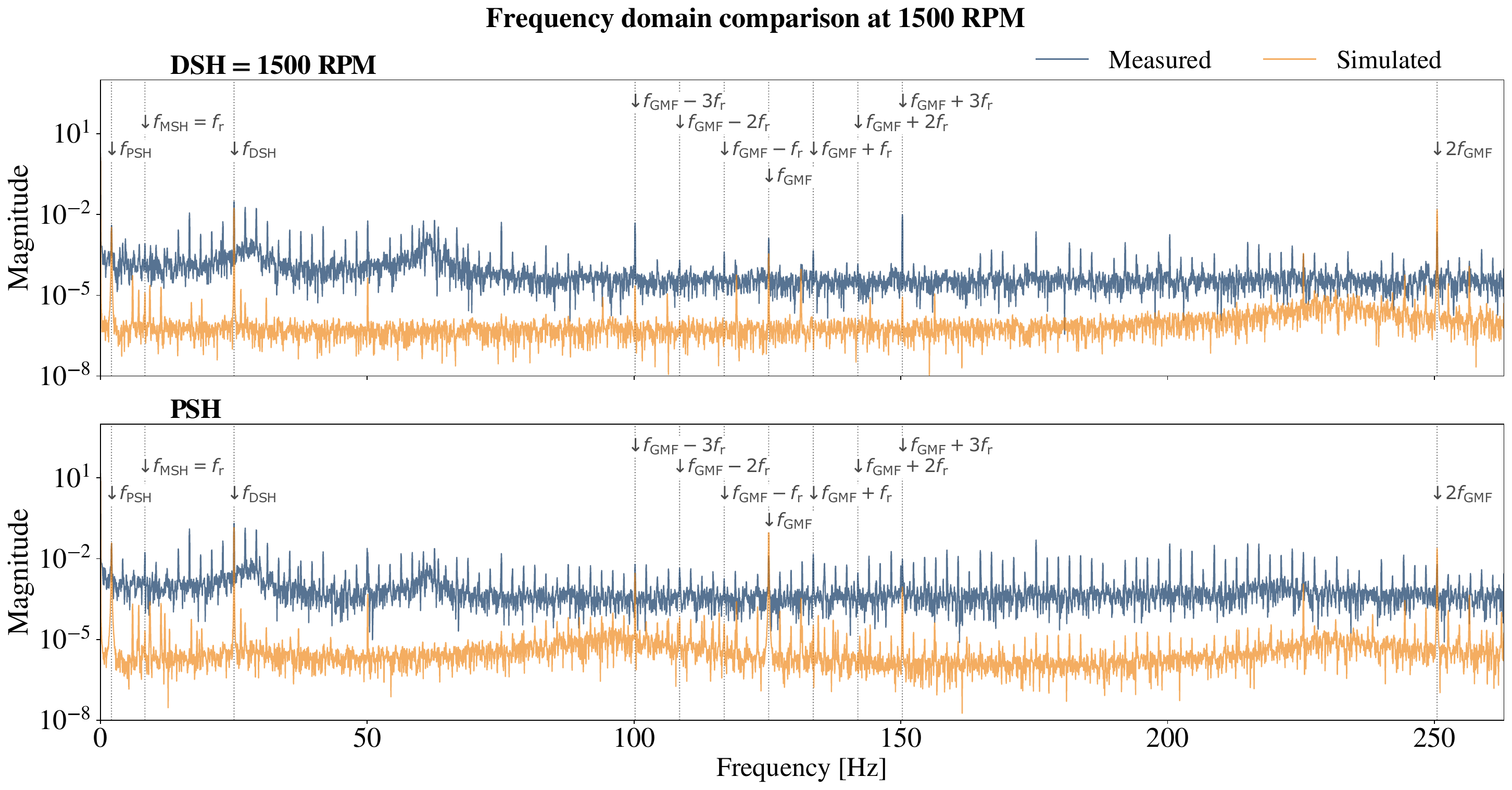} \vspace{-7mm}
        \caption{\label{fig:fft_healthy_1500rpm} DSH and PSH torque spectra with a healthy pinion at 1500 rpm drive shaft speed and 1.31~Nm torque load.}
    \end{figure}

\begin{figure}[H]
        \centering
        \includegraphics[width=\linewidth, trim={0cm 0cm 0cm 1.4cm},clip]{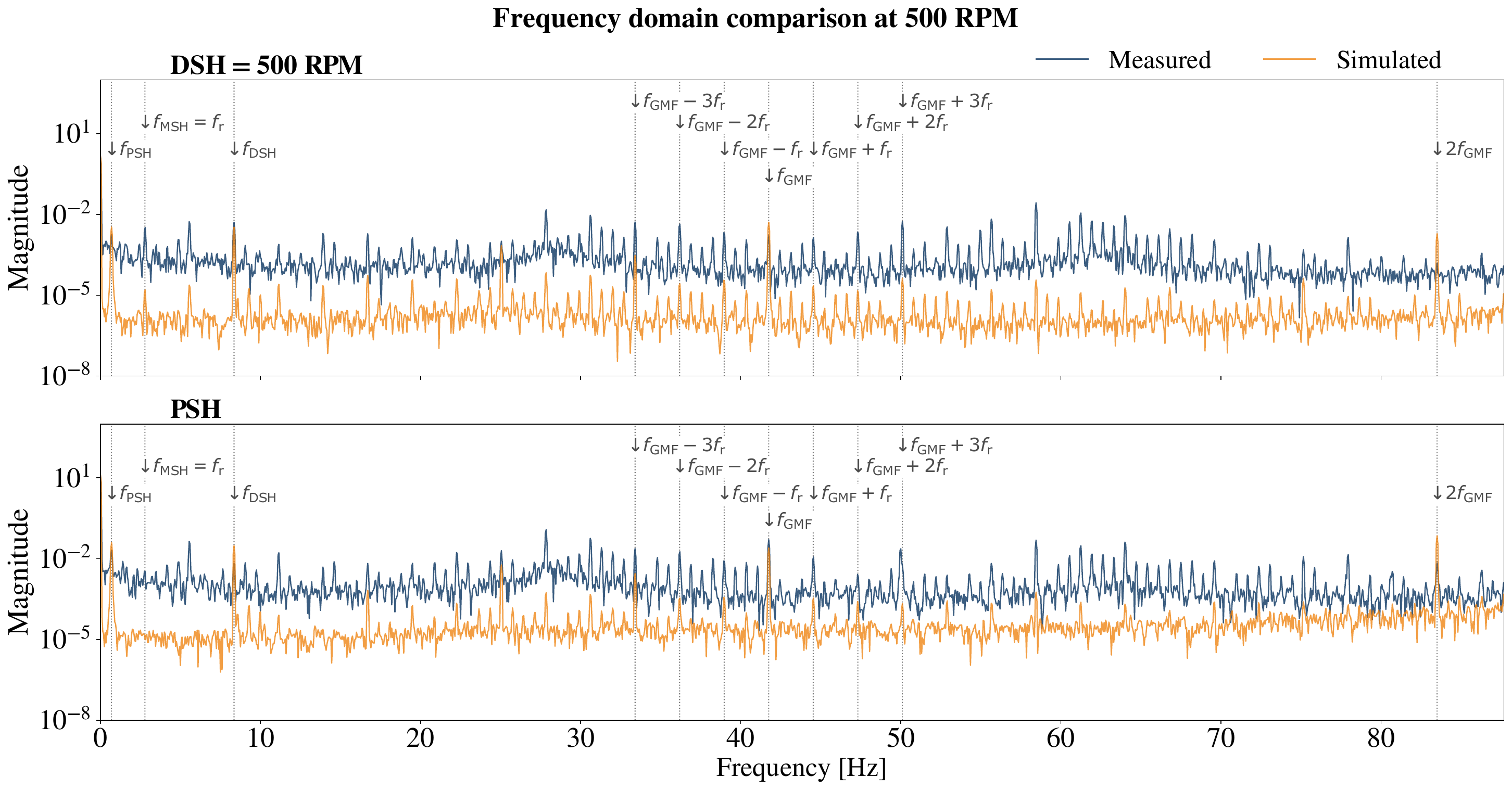} \vspace{-7mm}
        \caption{\label{fig:fft_mild_tff_500rpm} DSH and PSH torque spectra with a mild pinion TFF at 500 rpm drive shaft speed and 1.31~Nm torque load.}
    \end{figure}

\begin{figure}[H]
        \centering
        \includegraphics[width=\linewidth, trim={0cm 0cm 0cm 1.4cm},clip]{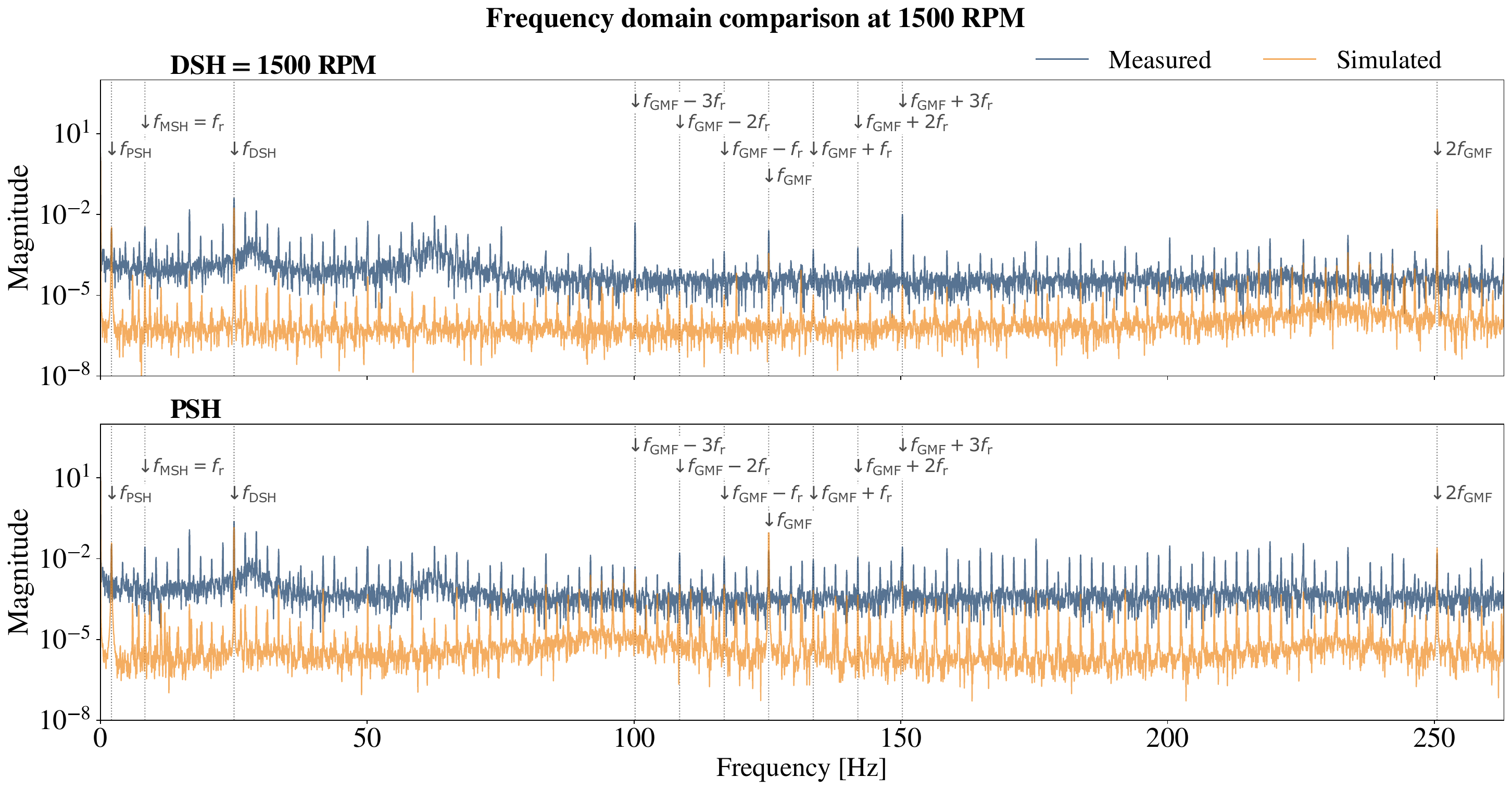} \vspace{-7mm}
        \caption{\label{fig:fft_mild_tff_1500rpm} DSH and PSH torque spectra with a mild pinion TFF at 1500 rpm drive shaft speed and 1.31~Nm torque load.}
    \end{figure}

\begin{figure}[H]
        \centering
        \includegraphics[width=\linewidth, trim={0cm 0cm 0cm 1.4cm},clip]{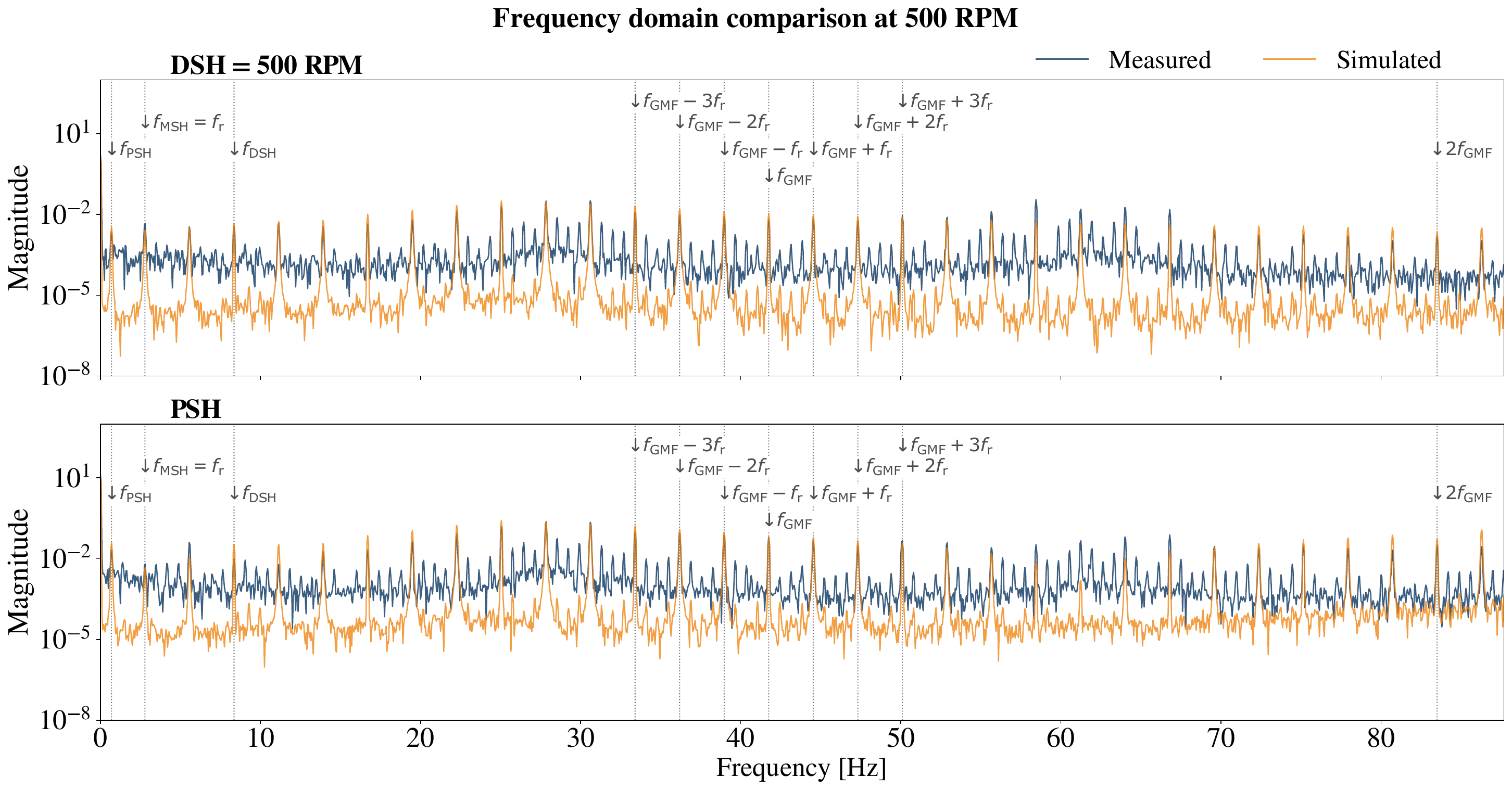} \vspace{-7mm}
        \caption{\label{fig:fft_severe_tff_500rpm} DSH and PSH torque spectra with a severe pinion TFF at 500 rpm drive shaft speed and 1.31~Nm torque load.}
    \end{figure}

\begin{figure}[H]
        \centering
        \includegraphics[width=\linewidth, trim={0cm 0cm 0cm 1.4cm},clip]{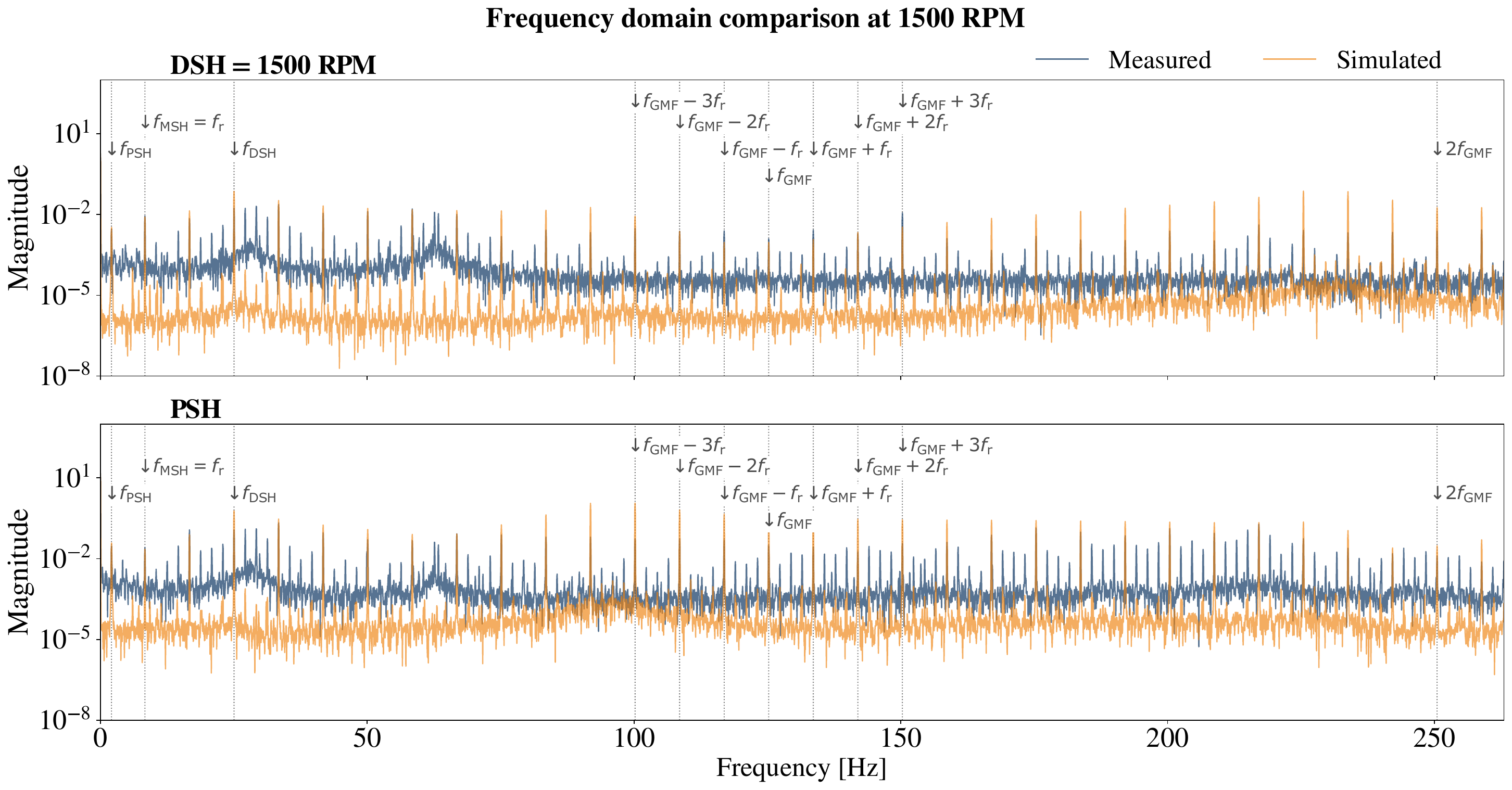} \vspace{-7mm}
        \caption{\label{fig:fft_severe_tff_1500rpm} DSH and PSH torque spectra with a severe pinion TFF at 1500 rpm drive shaft speed and 1.31~Nm torque load.}
    \end{figure}

    Measured and simulated torque signals were transformed to the frequency domain by using FFT analysis. The resulting frequency spectra at the representative speeds are presented in Figures \ref{fig:fft_healthy_500rpm}--\ref{fig:fft_severe_tff_1500rpm}. The vertical dashed lines in the figures mark the key frequencies calculated with Equations \eqref{eq:f_{GMF}} -- \eqref{eq:f_{sb}}. Rotational frequencies of the drive, middle and propeller shafts are denoted by $f_{DSH}$, $f_{MSH}$ and $f_{PSH}$, respectively. Gear mesh frequency and its second harmonic are $f_{GMF}$ and $f_{2GMF}$,  and associated sideband frequencies are denoted by $f_{GMF}\pm nf_r$. Numerical values for these frequencies at the representative speeds are listed in Table \ref{tab:freqs}.

    \begin{table}[width=0.6\linewidth,cols=5,pos=h!]
    \caption{Numerical values of the main frequencies expected to excite at the representative speeds.}
    \label{tab:freqs}
    \centering
    \begin{tabular*}{\tblwidth}{@{} CCCCC@{} }
        \toprule
         \textbf{Frequency}  & \textbf{250 RPM}    & \textbf{500 RPM}   & \textbf{1000 RPM}  & \textbf{1500 RPM}   \\ 
         \midrule
         $f_{PSH}$  &  1.39 Hz   & 0.70 Hz   & 5.57 Hz   & 2.09 Hz   \\
         $f_{MSH}$  &  4.18 Hz   & 2.78 Hz   & 16.70 Hz  & 8.35 Hz   \\
         $f_{DSH}$  &  16.70 Hz   & 8.35 Hz   & 66.80 Hz  & 25.05 Hz  \\
         $f_{GMF}$  &  20.86 Hz  & 41.75 Hz  & 83.50 Hz  & 125.25 Hz \\ 
         \bottomrule
    \end{tabular*}
\end{table}

    In the healthy torque spectra (Figures \ref{fig:fft_healthy_500rpm} and \ref{fig:fft_healthy_1500rpm}), both measured and simulated results show distinct peaks at the gear meshing frequency and its harmonics, reflecting smooth and periodic torsional vibrations. The most evident differences include a slightly higher broadband noise floor in the measured spectra, attributed to real-world conditions and structural resonances. Additionally, the measured spectra consistently display two broadband humps near 27 Hz and 60 Hz, absent in the simulations. The 27 Hz peak likely corresponds to a system resonance that remains unexcited in the simulation model, while the 60 Hz component may originate from unmodelled sources such as motor electromagnetics or simplified gear representations. Despite these discrepancies, overall spectral shapes and dominant peak amplitudes align well, supporting the capability to replicate frequency response of the system. Clear peaks are observed at $f_{PSH}$ and $f_{DSH}$, consistent with the torque sensor locations. The $f_{MSH}$ component only appears in the drive shaft spectra and at higher speeds. As expected under healthy conditions, sideband frequencies remain insignificant. Minor random spikes observed in the measured data are likely due to slight misalignments or manufacturing tolerances. An exception are the third sidebands around the main meshing frequency, where the peaks are from harmonics of the middle shaft rotational frequency. At higher speeds, noise levels increase modestly, particularly in the propeller shaft spectra, but still, the principal frequency components remain clearly distinguishable.

     \vspace{-5mm}
    
    In mild TFF conditions (Figures \ref{fig:fft_mild_tff_500rpm} and \ref{fig:fft_mild_tff_1500rpm}), measured and simulated spectra remain comparable regarding main frequency peaks, and small sideband spikes become noticeable, indicating an initial defect. Increased broadband noise, particularly in drive shaft spectra, further confirms the presence of a mild fault. Differences between healthy and mild TFF conditions are more pronounced in simulated than in measured spectra, possibly because small defects in experimental setups are more readily damped by lubrication or centrifugal forces at higher rotational speeds. Consequently, measured healthy and mild TFF spectra can appear to be quite similar, complicating their differentiation.

     \vspace{-5mm}
    
    For severe TFF (Figures \ref{fig:fft_severe_tff_500rpm} and \ref{fig:fft_severe_tff_1500rpm}), the frequency analysis shows substantial peak amplitudes at all key frequencies, including the sidebands in both measured and simulated spectra, clearly implying fault-induced gear meshing. Simulations effectively replicate the presence and positions of sideband peaks but also overestimate their magnitudes, particularly at higher rotational speeds. Fault-related harmonics become prominent, superimposed on meshing frequency harmonics, and broadband noise notably elevates. This elevation is clearer in simulations, particularly at higher speeds, resulting in more comparable measured and simulated broadband levels. Moreover, the $f_{MSH}$ frequency becomes as visible as the other rotational frequency components, even at lower speeds. 

     \vspace{-5mm}
    
\begin{figure}[H]
        \centering
        \includegraphics[width=0.98\linewidth, trim={0.6cm 0.5cm 0.8cm 0.8cm},clip]{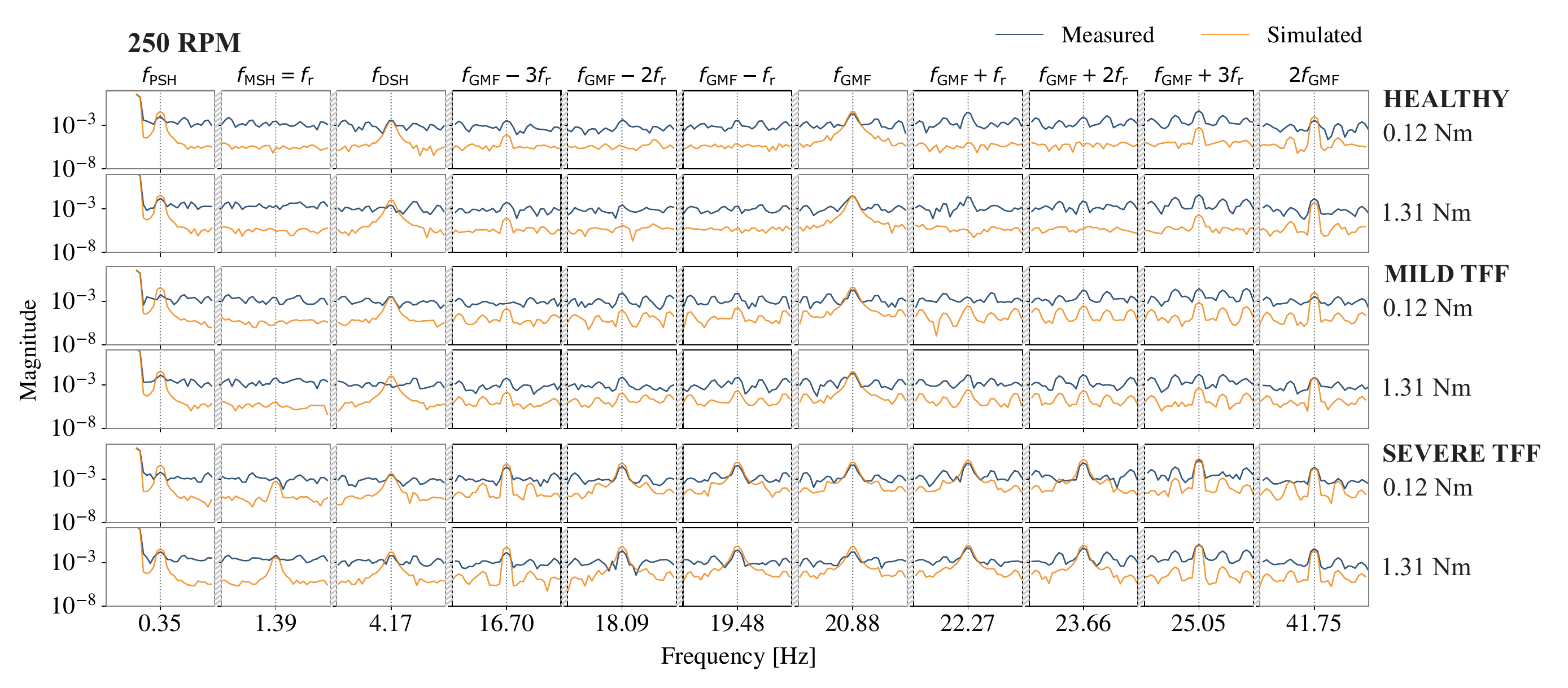}
        \caption{Torsional spectra at a drive shaft speed of 250 RPM for all health conditions and two torque loads, highlighting the key frequencies. \label{fig:250rpm_combined_spectra}}
    \end{figure}

\begin{figure}[H]
        \centering
        \includegraphics[width=0.98\linewidth, trim={0.6cm 0.5cm 0.8cm 0.6cm},clip]{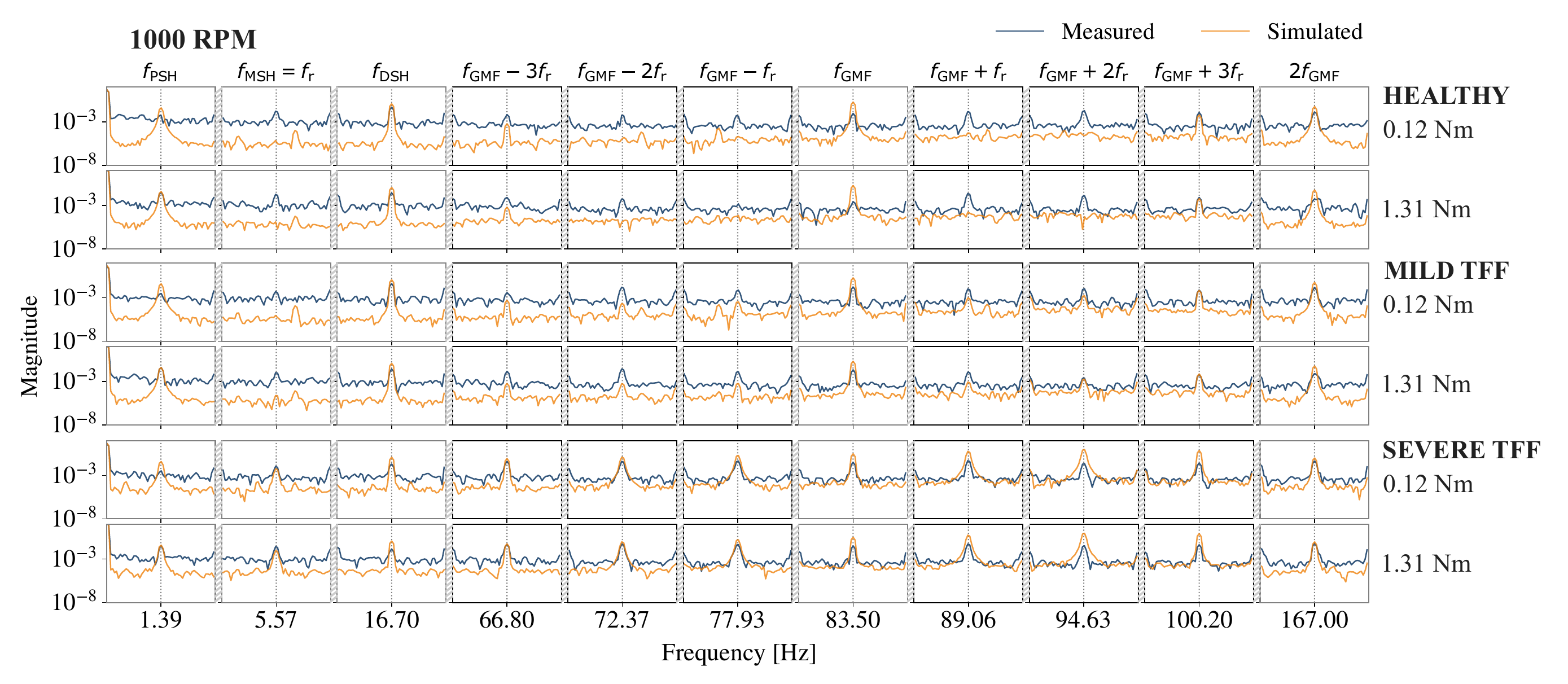}
        \caption{Torsional spectra at a drive shaft speed of 1000 RPM for all health conditions and two torque loads, highlighting the key frequencies \label{fig:1000rpm_combined_spectra}}
    \end{figure}

\begin{figure}[H]
        \centering
        \includegraphics[width=0.98\linewidth, trim={0.6cm 0.5cm 0.8cm 0.6cm},clip]{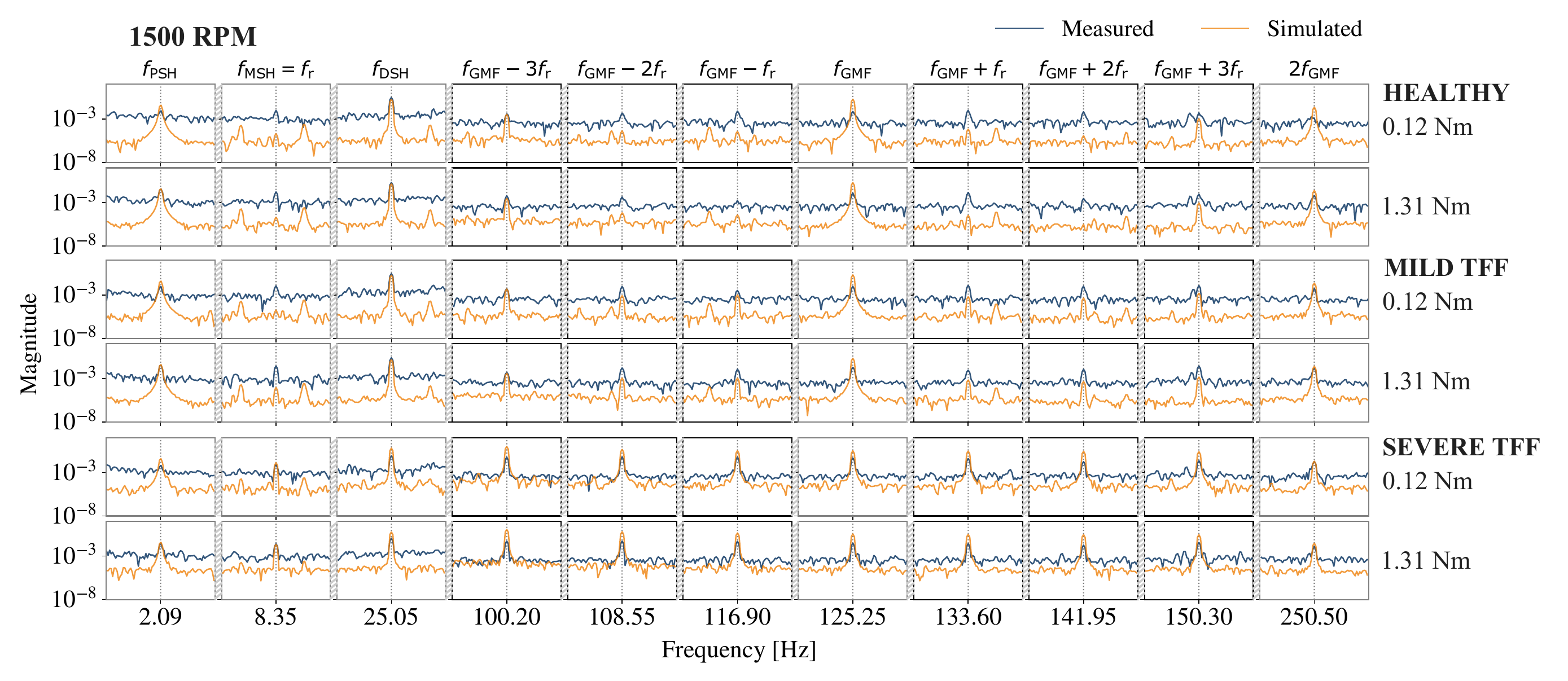}
        \caption{Torsional spectra at a drive shaft speed of 1500 RPM for all health conditions and two torque loads, highlighting the key frequencies \label{fig:1500rpm_combined_spectra}}
    \end{figure}

    Figures \ref{fig:250rpm_combined_spectra}--\ref{fig:1500rpm_combined_spectra} illustrate the frequency spectra for three selected speeds and two torque loads across all health conditions. The spectra are segmented into windows around key frequencies for clarity. These results demonstrate that drive speed has the greatest impact on peak amplitudes, while load variations primarily affect peak amplitudes at lower speeds.

\section{Future framework} \label{sec:sect_futurework}
The primary advantage of the proposed simulation framework is its ability to supplement and enrich measured datasets by generating fault scenarios and operating conditions that are too complex or otherwise impractical to realise experimentally. Laboratory datasets typically suffer from limitations such as restricted numbers of fault instances and operating points due to the inherent complexity and expense of physically inducing gear faults. For instance, existing datasets like UNSW TMCM \cite{chin2022}, MCC5-THU \cite{chen2024}, and the Aalto Gear Fault dataset \cite{dahl2024} provide only a limited number of fault variations and operating conditions.

By calibrating the PCM model primarily with existing healthy measurements, which are typically more available, the proposed simulation approach then facilitates the efficient and cost-effective generation of diverse fault geometries and severities, because the fault geometry can be adjusted just by directly modifying the component geometry in CAD software. This significantly enhances the variety of faulty data, improving the robustness and generalisability of data-driven diagnostic techniques, particularly in ML and DL applications used in PdM. In addition, simulated faults can be systematically varied across different operational speeds, loads, and other condition parameters, directly addressing dataset scarcity issues. By simulating the diverse variations of each fault type, the DL model learns to generalise to real-world faults, which are never identical.

Simulation data inherently involves critical dynamic features, such as gear mesh stiffness variations and fault-induced transient behaviour, which are essential for effective fault detection and diagnosis. While an inevitable simulation-to-reality gap exists, the inclusion of real-world measurements alongside simulation data in training datasets is anticipated to effectively mitigate this gap. Future research will investigate optimal ratios of real to simulated data in training sets and explore complementary methods, such as domain randomisation \cite{tobin2017, mehta2020, chen2021, karhinen2023} and generative adversarial networks (GANs) \cite{dimaggio2023}, to further enhance the utility and accuracy of simulated data in fault diagnosis and PdM frameworks.

\section{Conclusion} \label{sec:sect_conclusion}
This study developed and validated a comprehensive MBS model of a downscaled thruster drivetrain, focusing on accurate simulation of torsional bevel gear vibrations under both healthy and faulty conditions by using the PCM approach. The full drivetrain system, including multiple components and disturbance sources, was modelled to enable system-level analysis.

A key part of the validation was first calibrating the PCM parameters for a healthy bevel gearbox by comparing its contact properties to the well-established Hertzian-based Gear Pair element (FE225) within the complete powertrain model. This comparison was carried out using both simulation results and measured torsional vibrations from an azimuth thruster test rig.

After parameter validation under healthy conditions, the model was used to simulate mild and severe pinion tooth flank fractures. Corresponding torsional vibrations were analysed at two shaft locations not directly connected to the faulty gear. Model predictions were systematically compared with experimental measurements in both the time and frequency domains. The results successfully demonstrated that the PCM enabled realistic simulation of both healthy and faulty gear contacts, even when embedded with surrounding structures, yielding torsional vibrations closely comparable to measured signals.

In the time domain results, while deviations in amplitude and damping characteristics appeared, especially in the drive shaft at lower speeds, overall trends were well represented. The simulation accurately captured the relative increase in torque amplitude with rotation speed and matched the average torque levels with the measured data.

In the frequency-domain results, fault-related peak amplitudes appeared clearly at the expected frequencies, correctly reflecting higher amplitudes at increased speeds and fault severities. Even mild faults were distinguishable from the frequency spectra at appropriate speeds and loads. The measured spectra exhibited some additional features absent in the simulations, including a higher broadband noise floor and a greater number of frequency components. These differences are expected when comparing idealised simulation data with measurements from a noisy laboratory environment and are irrelevant for diagnostic purposes. 

This research demonstrates the strength of PCM in directly incorporating and varying realistic fault geometries in simulations, overcoming the constraints of conventional modelling and limited availability experimental data. By enabling the efficient generation of synthetic datasets with a wide range of fault types and operation conditions, the proposed approach addresses the data scarcity challenges inherent to data-driven PdM applications. Although the simulation provides valuable insights into system dynamics and gear fault impacts, accuracy regarding transient high-amplitude events and external disturbances could still be improved. Future work will focus on refining the system dynamics through more detailed modelling of damping, nonlinear characteristics, or lubrication, as well as systematically improving synthetic datasets by simulating diverse fault scenarios that are impractical to produce experimentally.

Moreover, future research will investigate the optimal balance between simulated and real measurement data in training sets to bridge the simulation-to-reality gap. Complementary methods such as domain randomisation and generative adversarial networks will also be explored to further improve the realism and utility of simulation-based datasets for robust and generalisable PdM solutions.

\bibliographystyle{model1-num-names} 
\bibliography{Bibliography/mssp_refs}

\end{document}